\documentclass[aps,superscriptaddress,showpacs,nofootinbib, 10pt]{revtex4-2}
\usepackage{mathrsfs}
\usepackage{bigints}
\usepackage{latexsym,bm}
\usepackage{graphicx}
\usepackage{indentfirst}
\usepackage{slashed}
\usepackage{amssymb}
\usepackage{amsmath}
\usepackage{bbm}
\usepackage{color}
\usepackage[dvipsnames]{xcolor}
\usepackage{epsfig}
\usepackage[titletoc]{appendix}
\usepackage{multirow}%
\usepackage{rotating}
\usepackage{epstopdf}
\usepackage{extarrows}
\usepackage{tabularx}
\usepackage{soul} 
\usepackage{xcolor}
\usepackage{lipsum}
\usepackage[colorlinks=true,allcolors=BlueViolet,hyperfootnotes=false]{hyperref}

\usepackage[utf8]{inputenc}

\newcommand{\be}{\begin{equation}} \newcommand{\ee}{\end{equation}}
\newcommand{\ba}{\begin{array}{c}} \newcommand{\ea}{\end{array}}
\newcommand{\bea}{\begin{eqnarray}} \newcommand{\eea}{\end{eqnarray}}
\newcommand{\beas}{\begin{eqnarray*}} \newcommand{\eeas}{\end{eqnarray*}}

\newcommand{\no}{\nonumber\\}

\begin{document}
\title{Tau polarization, new physics and CP violation in $\nu_\tau$–nucleus hyperon production}
\author{E. Hern\'andez}
\affiliation{Departamento de F\'\i sica Fundamental 
  e IUFFyM,\\ Universidad de Salamanca, E-37008 Salamanca, Spain}
\author{J. Nieves}
\affiliation{Instituto de F\'{\i}sica Corpuscular (centro mixto CSIC-UV), Institutos de Investigaci\'on de Paterna,
Apartado 22085, 46071, Valencia, Spain}
\author{J. E. Sobczyk}
\affiliation{Department of Physics, Chalmers University of Technology, 
SE-412 96 G\"oteborg, Sweden}

\date{\today}


\begin{abstract}
Charged-current interactions of the $\tau$ neutrino are the least constrained sector of the 
Standard Model (SM) flavour structure, and the polarisation of the produced $\tau$ lepton probes 
new-physics (NP) couplings, including CP-violating ones. 
Hyperon and charmed-baryon production gives access to the $u\to s$ and $d\to c$ transitions, 
but any measurement would be made on nuclei, which calls for a proper treatment of nuclear effects.
We therefore study the sequential reactions 
$\nu_\tau\,{}^A Z\to\tau^-(\to\pi^-\nu_\tau)\,\Lambda_c^+ X$ and 
$\bar\nu_\tau\,{}^A Z\to\tau^+(\to\pi^+\bar\nu_\tau)\,\Lambda X$ within the impulse 
approximation. The energy and  angular distributions of the pion, which encode the $\tau$ 
polarization, are derived for an  arbitrary nuclear hole spectral function. As for free nucleons, 
an azimuthal pion asymmetry with respect to 
the plane spanned by the neutrino and the outgoing baryon  is shown to remain. This is a CP-odd  
observable that vanishes in the SM, so that a nonzero value  would signal CP violation irrespective of 
the hadronic form factors. Local Fermi gas results  for ${}^{16}$O show that nuclear effects lower the 
reaction thresholds but leave the pion  distributions, and hence the polarization information, close
to their free-nucleon form. For  representative NP scenarios the predicted asymmetries are at the 
percent level or below, with  a magnitude that, in the $\Lambda$ channel, depends strongly on the 
transition form factors.
\end{abstract}

\maketitle

\section{Introduction}
\label{sec:intro}

The charged-current (CC) interactions of the $\tau$ neutrino are the least
tested part of the flavour structure of the Standard Model (SM). The number of
$\nu_\tau$ CC events identified to date remains small, and comes from only a
handful of experiments~\cite{DONUT:2000fbd,OPERA:2015wbl,IceCube:2019dqi,FASER:2023zcr,SNDLHC:2023pun},
so that the $\nu_\tau$ sector is constrained far more weakly than its electron and muon
counterparts. At the same time persistent hints
of lepton-flavour-universality (LFU) violation in $b\to c\,\tau\bar\nu_\tau$
transitions~\cite{HFLAV:2022pwe} keep pointing towards new physics (NP) coupled
preferentially to the third generation. Several future facilities will
improve this situation substantially: the $\nu_\tau$ appearance program at
DUNE~\cite{DUNE:2020ypp}, beam-dump experiments such as SHiP~\cite{SHiP:2021nfo,Climescu:2026omx},
and the LHC forward physics program~\cite{FASER:2023zcr,SNDLHC:2023pun,DsTau:2019wjb}. It is therefore
timely to identify observables in $\nu_\tau$-induced reactions that are
sensitive to NP and that can be extracted from these measurements.

Below the electroweak scale, NP contributions to semileptonic transitions can be
systematically described by an effective Hamiltonian comprising the complete set of dimension-six
four-fermion operators~\cite{Cirigliano:2012ab,Bhattacharya:2011qm}. For the light-quark 
transitions, the electron- and muon-flavoured Wilson coefficients are tightly bounded by 
super-allowed $\beta$ decays, by pion and kaon leptonic and semileptonic decays, and by 
Cabibbo-Kobayashi-Maskawa (CKM) unitarity~\cite{Gonzalez-Alonso:2018omy}, whereas the 
$\tau$-flavoured ones are much less constrained, being probed almost exclusively through hadronic 
$\tau$ decays~\cite{Cirigliano:2018dyk}. For $d\to c$ transitions, the available constraints arise
only indirectly from charmed-meson decays~\cite{Leng:2020fei}. Direct measurements using $\tau$ (anti)neutrino
beams therefore offer a unique opportunity to explore a largely unconstrained region of parameter space.

Among the possible reactions, the associated production of a hyperon or a charmed baryon,
$\bar\nu_\tau p\to\tau^+\Lambda$ and $\nu_\tau n\to\tau^-\Lambda_c^+$, is particularly attractive.
First, these reactions probe quark transitions, $u\to s$ and $d\to c$, that are absent from the
traditional quasielastic (QE) channel involving only nucleons. Second, the final-state baryon is
readily distinguishable from a knocked-out nucleon and produces visible decay signatures,
facilitating event identification. The price to pay is a higher energy threshold than for QE
$\tau$ production, so that these reactions are accessible primarily in the high-energy tails of
accelerator neutrino fluxes and in atmospheric-neutrino and beam-dump environments.

The $\tau$ polarisation provides information beyond that contained in the event rate. Its
longitudinal component is sensitive to combinations of NP couplings that can leave the integrated
cross section essentially unchanged, helping to resolve degeneracies that rate measurements alone
cannot lift, while the transverse component normal to the scattering plane is odd under time
reversal and therefore sensitive to CP violation. Because the $\tau$ lepton decays before it can be
tracked, its polarisation must be inferred from the energy and angular distributions of its decay
products, the two-body decay $\tau\to\pi\nu_\tau$ (${\cal B}\simeq 10.8\%$~\cite{ParticleDataGroup:2026mpi})
being the cleanest channel. In a
previous work~\cite{Hernandez:2026svn}, we developed this framework for the free-nucleon reactions
$\nu_\tau n \to \Lambda_c^+\tau^-(\pi^-\nu_\tau)$ and $\bar\nu_\tau p \to \Lambda\tau^+(\pi^+
\bar\nu_\tau)$, and showed that the asymmetry in the number of pions emitted above and below the
plane spanned by the incoming neutrino and the outgoing baryon is directly proportional to the
CP-violating NP contributions to the underlying quark-level transition. It thus vanishes in the SM
and provides a CP-odd observable with no background.

The process $\nu_\tau n\to\tau^-\Lambda_c$, including NP effects in the $d\rightarrow c$ transition,
was first studied in Ref.~\cite{Kong:2023kkd}, where the polarisation components of the $\tau$
lepton were also presented.\footnote{The SM cross sections reported in Ref.~\cite{Kong:2023kkd} are
a factor of two larger than those obtained here and in Refs.~\cite{Hernandez:2025snr,Hernandez:2026svn}
at the nucleon level, owing to the omission of the initial-nucleon spin-averaging
factor~\cite{sxyan2026private}. Moreover, as argued in Ref.~\cite{Hernandez:2025snr}, the allowed
ranges for the imaginary parts of some Wilson coefficients reported in Ref.~\cite{Kong:2023kkd} are
inconsistent with the constraints derived from high-$p_T$ lepton tails in Drell--Yan
data~\cite{Fuentes-Martin:2020lea}. Similar considerations apply to the study of
$e^-p\to\Lambda_c^+\tau^-$ in Ref.~\cite{Yan:2024bce}, which in addition appears to consider only a
left-handed incoming electron; for an unpolarised electron beam, the cross section should include an
additional factor of $1/2$.} These components can be partially inferred from the visible decay 
products, as demonstrated in Ref.~\cite{Hernandez:2026svn} for a free-neutron target and in the 
present work for nuclear targets. The analyses of Refs.~\cite{Kong:2023kkd,Hernandez:2026svn}, 
however, were performed at the nucleon level, whereas all the neutrino experiments mentioned above use nuclear targets
--argon, water and ice, iron, or the heavy nuclei in emulsion-- and no free-neutron target is
available. Extracting NP constraints from these processes therefore requires a nuclear calculation.
As discussed in Ref.~\cite{Hernandez:2025snr}, nuclear
effects are likely to play a significant role, and the associated uncertainties may hinder the
extraction of reliable constraints on possible NP Wilson coefficients, even under optimistic
assumptions such as perfect knowledge of the nucleon form factors and extremely high event statistics.

The nuclear medium smears the initial nucleon momentum and modifies the energy balance through
binding effects, redistributing the reaction strength and effectively lowering the threshold. In
addition, the outgoing baryon interacts with the residual nucleus, further modifying the final state.
There is, however, a further, less obvious consequence for the polarization observables themselves.
For a free nucleon, the $\tau$ momentum, and hence the axes with respect to which the polarization
components are defined, can be reconstructed event by event. This is no longer possible for a
nuclear target, where the initial nucleon momentum is unobserved, raising the question of how much 
of their sensitivity to NP survives the resulting kinematic smearing. 

To address this question, we derive, within the impulse approximation, general expressions for the 
sequential reactions $\nu_\tau\,^A Z \to \tau^-(\pi^-\nu_\tau)\,Y^+ X$ and
$\bar\nu_\tau\,^A Z \to \tau^+(\pi^+\bar\nu_\tau)\,Y^0 X$,
where $Y$ denotes a hyperon or charmed baryon, and $X$ the residual 
nuclear system containing $(A-1)$ nucleons. The formalism accommodates an arbitrary neutrino 
spin-density matrix and all dimension-six operators, including those involving right-handed 
neutrino fields, although in the numerical applications we assume SM-like fully polarized beams. 
We apply it to $\Lambda$ and $\Lambda_c$ production and compute the energy and angular 
distributions of the charged pion emitted in the $\tau$ decay, which retain the information on the 
$\tau$ polarization after averaging over the nuclear initial state. Finally, we show that the CP-odd 
azimuthal pion asymmetry of Ref.~\cite{Hernandez:2026svn} remains well defined for nuclear targets, 
even though Fermi motion allows the intermediate $\tau$ to emerge outside the reference plane.

The paper is organised as follows. In Sec.~\ref{sec:nuclear} we derive the cross section for the
$\nu_\tau\,n \to \Lambda_c\,\tau^-(\pi^-\nu_\tau)$ reaction in a nucleus, introducing the nuclear
tensors and the $\tau$ polarisation vector averaged over the initial state. Section~\ref{sec:CP-asy}
defines the azimuthal pion asymmetry, shows how it is related to CP-violating NP contributions, and in Sec.~\ref{sec:anti} we extend 
the formalism to the antineutrino-induced $\bar\nu_\tau\,p \to \Lambda\,\tau^+(\pi^+\bar\nu_\tau)$ 
reaction. In Sec.~\ref{sec:LFG} we discuss the treatment of nuclear effects and the local Fermi gas 
(LFG) model used in this work to obtain numerical results. Numerical results for $\Lambda_c$
and $\Lambda$ production off ${}^{16}$O, including the comparison with the free-nucleon case, the
role of the $\tau$ polarisation in the pion distributions and the CP asymmetry in representative NP
scenarios, are discussed in Sec.~\ref{sec:results}. Finally, the main
conclusions are summarised in Sec.~\ref{sec:conclusions}. Technical details are addressed in the
appendices: App.~\ref{app:details} collects the phase-space integrations, the integration limits
for $E_\pi$ and $\cos\theta_\pi$, and the specific LFG results for the cross section, while
App.~\ref{app:asy} gives the details of the derivation of the CP asymmetry, including its LFG expression.

%
%
%
\section{The  $\nu_\tau n\to
\Lambda_c\tau^-(\pi^-\nu_\tau $) reaction in a nucleus}
\label{sec:nuclear}
As in Ref.~\cite{Hernandez:2026svn}, we work with the low-energy effective Hamiltonian that 
includes all dimension-six operators contributing to $q\to q'\ell\nu_\ell$ transitions for 
$d\to u,c$ and $u\to d,s$, allowing for both left- and right-handed neutrino 
fields~\cite{Cirigliano:2012ab,Bhattacharya:2011qm}.
To evaluate nuclear effects in the $\nu_\tau n\to\Lambda_c\tau^-(\pi^-\nu_\tau)$ reaction we start from  Eq.~(11) in 
Ref.~\cite{Hernandez:2026svn} where the corresponding cross section for the reaction at the 
nucleon level is given as
\bea
\sigma
&=&\frac{ {\cal B}(\tau\to\pi\nu_\tau)m^2_\tau M_{\Lambda_c}G_F^2|V_{cd}|^2}
{4\pi^3|\vec k\,|(m^2_\tau-m^2_\pi)}
\int\frac{d^3p'}{E'}\int\frac{d^3p_\pi}{ E_\pi}\int\frac{d^3p_{\nu_\tau}}
{ E_{\nu_\tau}}
\delta^{(4)}(k+p-p'-p_\pi-p_{\nu_\tau})
\delta((k+p-p')^2-m^2_\tau)\nonumber\\
&\times&\Big[\sum_{\chi=L,R}\rho_{\chi\chi}\sum_{a,b}W^{ab}_\chi L_{\chi ab}\Big]
\Big(1+\frac{2m_\tau}{m^2_\tau-m^2_\pi}\,p_\pi\cdot{\cal P}
\Big)
\label{eq:WLsigma}
\eea
where $k$, $p$, $p'$, $p_\pi$, and $p_{\nu_\tau}$ denote the four-momenta of the incoming $\nu_\tau$,
the initial nucleon, the final-state hyperon, the pion, and the $\nu_\tau$ produced in the $\tau$ 
decay, respectively. The Fermi constant is denoted by $G_F$, and $V_{cd}$ is the relevant CKM matrix 
element. The quantities $\rho_{LL}$ and $\rho_{RR}$ denote the diagonal elements of the incident 
neutrino density matrix.\footnote{The off-diagonal contributions vanish when neutrino mass terms are 
neglected.} We also denote by ${\cal B}(\tau\to\pi\nu_\tau)$ the branching fraction for the 
$\tau\to\pi\nu_\tau$ decay. This decay mode is particularly clean for polarimetry, as it is a 
two-body process and the pion provides maximal spin-analysing power. For massless neutrinos, or at energies much larger than their mass, 
chirality and helicity coincide: left- and right-handed chirality correspond to negative and positive
helicity, respectively. We adopt this identification throughout. Even in the presence of NP, however,
neutrino production is expected to be dominated by SM interactions. Accordingly, as anticipated in 
the Introduction, we assume fully left-handed neutrino beams ($\rho_{LL}=1$, $\rho_{RR}=0$) and 
fully right-handed antineutrino beams ($\rho_{LL}=0$, $\rho_{RR}=1$). Consequently, operators 
involving right-handed neutrino fields do not contribute at the primary vertex.\footnote{This is 
the conservative choice. A right-handed neutrino coupling large enough to be relevant here would 
generally also induce a wrong-helicity component in the neutrino beam produced through for instance 
$D_s\to\tau\nu_\tau$. Setting $\rho_{RR}=0$ therefore amounts to assuming that beam production is 
dominated by the SM.} On the other hand, $W^{ab}_\chi$ and 
$L_{\chi ab}$ denote the hadronic and leptonic tensors, respectively, where the indices $a$ and $b$
label the three allowed types of Dirac structures: (pseudo-)vector, (pseudo-)scalar, and 
(pseudo-)tensor. The four-vector $\cal P^\mu$ describes the $\tau$ polarisation. The hadronic
tensors are constructed from the hadron masses and four-momenta and expressed in terms of 
dimensionless structure functions $\widetilde W_\chi$, which depend on the form factors and the 
Wilson coefficients of the effective Hamiltonian. A comprehensive account of both $W^{ab}_\chi$ and
$L_{\chi ab}$ can be found in Ref.~\cite{Hernandez:2026svn} and references therein. In particular, 
the leptonic tensors are obtained from those in Appendix B of Ref.~\cite{Penalva:2021wye} by 
multiplying the unpolarised contribution by a factor of two.

For  convenience in what follows, we repeat here a few general expressions given in 
Ref.~\cite{Hernandez:2026svn}. We have that
\bea
\sum_{\chi=L,R}\rho_{\chi\chi}\sum_{a,b}W^{ab}_\chi L_{\chi ab}=M^2_n\widetilde{\cal N}(\omega,k\cdot p)=
\frac{M^2_n}{2}\Big[-{\cal A}(\omega)+{\cal B}(\omega)
\frac{p\cdot k}{M^2_n}-{\cal C}(\omega)
\frac{(p\cdot k)^2}{M^4_n}\Big], 
\eea
while ${\cal P}^\mu$ is given by
\bea
{\cal P}^\mu=\frac 1{\widetilde{\cal N}(\omega,k\cdot p)}
\Big[\frac{p^\mu_\perp}{M_n}\widetilde{\cal N}_{{\cal H}1}(\omega,k\cdot p)+
\frac{q^\mu_\perp}{M_n}\widetilde{\cal N}_{{\cal H}2}(\omega,k\cdot p)
+\frac{\epsilon^{\mu k'qp}}{M^3_n}\widetilde{\cal N}_{{\cal H}3}(\omega,k\cdot p)\Big]
\eea
where $q=p-p'$, $\epsilon^{\mu k' q p }= \epsilon^{\mu \lambda \nu \rho} k'_\lambda q_\nu p_\rho $ 
and $\epsilon_{0123}=+1$, $\ell_\perp=\ell-\frac{\ell\cdot k'}{m_\tau^2}k'$ and $k'=q+k$ is the 
intermediate $\tau$ four-momentum. Besides,
\bea
\widetilde{\cal N}_{{\cal H}1}(\omega,k\cdot p)&=&-{\cal A}_{{\cal
H}}(\omega)+{\cal C}_{{\cal
H}}(\omega)
\frac{p\cdot k}{M^2_n} \label{eq:pol1},\\
\widetilde{\cal N}_{{\cal H}2}(\omega,k\cdot p)&=&-{\cal B}_{{\cal
H}}(\omega)+{\cal D}_{{\cal
H}}(\omega)
\frac{p\cdot k}{M^2_n}-{\cal E}_{{\cal
H}}(\omega)
\frac{(p\cdot k)^2}{M^4_n},\label{eq:pol2}\\
\widetilde{\cal N}_{{\cal H}3}(\omega,k\cdot p)&=&-{\cal F}_{{\cal
H}}(\omega)+{\cal G}_{{\cal
H}}(\omega)
\frac{p\cdot k}{M^2_n},\label{eq:nh3}
\eea
 Here, $\omega$ is the product of the neutron and $\Lambda_c^+$ four-velocities and is related to 
 the squared momentum transfer through $q^2=M_n^2+M_{\Lambda_c}^2-2M_nM_{\Lambda_c}\,\omega$.
The expressions for the functions ${\cal A}(\omega)$, ${\cal B}(\omega)$, and ${\cal C}(\omega)$  and
for ${\cal A}_{\cal H}(\omega)$, ${\cal B}_{\cal H}(\omega)$, ${\cal C}_{\cal H}(\omega)$, ${\cal D}
_{\cal H}(\omega)$, ${\cal E}_{\cal H}(\omega)$, ${\cal F}_{\cal H}(\omega)$, and ${\cal G}_{\cal H}
(\omega)$, in terms of the $\widetilde W_\chi$ structure functions are given in Appendix~D of 
Ref.~\cite{Penalva:2021wye}. In that reference, the quantities $M$, $M_\omega$, and $m_\ell$ 
correspond to $M_n$, $(M_n - M_{\Lambda_c}\omega)$, and $m_\tau$, respectively. There, these 
functions are obtained by summing over $\chi = L, R$, as the analysis involved a final-state neutrino
with a small but non-vanishing mass, allowing both chiralities, while neutrino-mass–suppressed terms
as, for instance, left–right interference terms, were neglected. In the present case, the neutrino is
incoming, and for a purely left-handed neutrino (right-handed antineutrino) beam produced by SM 
interactions, only the $\chi = L$ ($\chi = R$) contributions should be retained. More generally, the
factor $\rho_{\chi\chi}$ must be included in the chirality sums entering the definition of these 
functions. It is also worth mentioning that  both ${\cal F}_{{\cal H}}(\omega)$ and 
${\cal G}_{{\cal H}}(\omega)$ are nonzero only in the presence of CP-violating NP beyond the standard
model~\cite{Penalva:2021gef}.

As usual, ${\cal P}^\mu$ can be further expanded as
\bea
{\cal P}^\mu={\cal P}_L N^\mu_L+{\cal P}_T N^\mu_T+{\cal P}_{TT} N^\mu_{TT}\label{eq:PLTTT}
\eea
where the three space-like orthogonal four-vectors $N^\mu_{L,T,TT}$ are defined as
\bea
N^\mu_L=\Big(\frac{|\vec k\,'|}{m_\tau},\frac{k^{\prime0}\,\vec k\,'}
{m_\tau|\vec k\,'|}\Big),\ N^\mu_T=\Big(0,\frac{(\vec k\times\vec k\,')
\times\vec k\,'}
{|(\vec k\times\vec k\,')\times\vec k\,'|}\Big),\ 
N^\mu_{TT}=\Big(0,\frac{\vec k\times\vec k\,'}
{|\vec k\times\vec k\,'|}\Big)
\eea
and where ${\cal P}_a=-\left(N_a\cdot {\cal P}\right)$, with $a=L,T,TT$.
The longitudinal component ${\cal P}_L$ probes the helicity composition of the produced $\tau$ and 
is sensitive to combinations of scalar, pseudoscalar and tensor couplings, as well as right-handed 
quark currents. The transverse components probe the interference between operators of different 
chiralities: ${\cal P}_T$, which lies in the lepton scattering plane, is sensitive to the real parts
of the Wilson coefficients, whereas ${\cal P}_{TT}$, normal to this plane, is odd under time reversal
and probes their relative phases.

So far, we have presented the expression for the reaction at the nucleon level. To obtain the 
corresponding result for a nuclear target $A_Z$ with $Z$ protons, we must sum over the $(A-Z)$ 
neutrons in the nucleus. This involves introducing the appropriate dimensionless integration
\bea
f(E,\vec p\,)\to \int\frac{d^3p}{(2\pi)^3}\int dE\, S_h^{(n)}(E,\vec p\,)f(E,\vec p\,), \qquad \qquad \int\frac{d^3p}{(2\pi)^3}\int d E\, S_h^{(n)}( E,\vec p\,)=A-Z
\eea
where $S_h^{(n)}(E,\vec p\,)$ is the neutron-hole spectral function (SF), which gives the probability
density of finding a neutron in the target nucleus with momentum $\vec p$ and energy $E$. For a 
nucleus at rest, rotational invariance implies that the SF depends only on the magnitude of the 
neutron momentum, i.e., $S_h^{(n)}(E,\vec p\,)=S_h^{(n)}(E,|\vec p\,|)$. Therefore, we obtain

\bea
\sigma_{A_Z}
&=&\frac{{\cal B}(\tau\to\pi\nu_\tau) m^2_\tau G_F^2|V_{cd}|^2}
{4\pi^3|\vec k\,|(m^2_\tau-m^2_\pi)}
\int\frac{d^3p}{(2\pi)^3}\int dE\frac{M_n}{E}S_h^{(n)}(E,|\vec p\,|) 
\int{d^3p'}\frac{M_{\Lambda_c}}{E'}\int\frac{d^3p_\pi}{ E_\pi}
\int\frac{d^3p_{\nu_\tau}}{ E_{\nu_\tau}}\no
&\times&
\delta^{(4)}(k+p-p'-p_\pi-p_{\nu_\tau})
\delta\left((k+p-p')^2-m^2_\tau\right)\Big[\sum_{\chi=L,R}\rho_{\chi\chi}\sum_{a,b}W^{ab}_\chi L_{\chi ab}\Big]\Big(1+\frac{2m_\tau}{m^2_\tau-m^2_\pi}\,p_\pi\cdot{\cal P}
\Big). \label{eq:sigmaAZ1}
\eea
The $M_n/E$ factor included in the above equation accounts for the covariant normalization of the 
Dirac-spinor of the initial nucleon when it is not at rest, while the on-shell energies of the initial 
nucleon and the final hadron are $\sqrt{M_n^2+\vec p\,^2}$ and $E'=\sqrt{M_{\Lambda_c}^2+(\vec p-\vec
q\,)^2}$, respectively. At first sight, using on-shell energies  amounts to neglecting the effects of the 
nuclear medium on the produced charmed hyperon. However, this interpretation is not entirely 
accurate, as we discuss in Sec.~\ref{sec:LFG}.

Introducing in Eq.~\eqref{eq:sigmaAZ1} the identity\footnote{ By introducing $k'$, we can rewrite 
$q^2$ as $(k'-k)^2$, making it independent of the initial nucleon four-momentum $p^\mu$. 
Consequently, the transition form factors can be taken outside the $d^3p$ and $dE$ integrals,
considerably simplifying the calculation.} $\int d^4k'\,\delta^{(4)}(p_\pi+p_{\nu_\tau}-k')=1$, 
one arrives at
\bea
\sigma_{A_Z}
&=&\frac{ {\cal B}(\tau\to\pi\nu_\tau) m^2_\tau G_F^2|V_{cd}|^2}
{8\pi^3|\vec k\,|(m^2_\tau-m^2_\pi)}
\int\frac{d^3k'}{k^{\prime0}}\int\frac{d^3p_\pi}{ E_\pi\,|\vec k\,'-\vec p_\pi|}
\,\delta(k^{\prime0}-E_\pi-|\vec k\,'-\vec p_\pi|)
\no
&\times&\int\frac{d^3p}{(2\pi)^3}\int dES_h^{(n)}(E,|\vec p\,|)
\frac{M_nM_{\Lambda_c}}{E\, E'}\,
 \delta(E-E'-q^0)
\Big[\sum_{\chi=L,R}\rho_{\chi\chi}\sum_{a,b}W^{ab}_\chi L_{\chi ab}\Big]
\Big(1+\frac{2m_\tau}{m^2_\tau-m^2_\pi}\,p_\pi\cdot{\cal P}
\Big),
\label{eq:sigaz}
\eea
with $ q=k'-k$ and  $k^{\prime0}=\sqrt{m^2_\tau+\vec k^{\,\prime2}}$.

We now introduce the nuclear tensors and the corresponding $\tau$-polarization vector for this 
reaction. The nuclear tensors are defined as
\bea
W_{A_Z; \chi}^{ab}(q)=\int\frac{d^3p}{(2\pi)^3}\int dES_h^{(n)}(E,|\vec p\,|)
\frac{M_nM_{\Lambda_c}}{E\, E'}\,
 \delta(E-E'-q^0)W^{ab}_\chi(p,q),
 \label{eq:nuctenquasie}
\eea
The present definition is consistent with those adopted in 
Refs.~\cite{Nieves:2004wx,Valverde:2006yi,Sobczyk:2019uej,Hernandez:2022nmp}, with the only 
difference that the corresponding $W_{A_Z; \chi}^{ab}$ tensors in those works include the squared 
CKM matrix element. In addition, the analyses of 
Refs.~\cite{Nieves:2004wx,Valverde:2006yi,Sobczyk:2019uej,Hernandez:2022nmp} are restricted to 
the SM tensors, i.e., those generated solely by the $V-A$ current.

Similarly, we define the polarization vector for this process as
\bea
{\cal P}^\mu_{A_Z}(q) &=&\frac{M_nM_{\Lambda_c}}{F^{A_Z}_{\nu_\tau \Lambda_c}(q)}
\int\frac{d^3p}{(2\pi)^3}\int \frac{dE}{E\, E'}S_h^{(n)}(E,|\vec p\,|)\,
 \delta(E-E'-q^0)\Big[\sum_{\chi=L,R}\rho_{\chi\chi}\sum_{a,b}W_\chi^{ab}L_{\chi\,ab}\Big]
 {\cal P}^\mu(p,q),
 \label{eq:polnuc}
\eea
where  we have introduced the nuclear spin-averaged squared amplitude $F^{A_Z}_{\nu_\tau \Lambda_c}$,
which has dimensions of energy in natural units and is given by
\be 
F^{A_Z}_{\nu_\tau \Lambda_c}(q) = \Big[\sum_{\chi=L,R}\rho_{\chi\chi}\sum_{a,b}
W_{A_Z;\chi}^{ab}L_{\chi\,ab}\Big].
\label{eq:faz}
\ee
The four-vector ${\cal P}_{A_Z}$ is manifestly orthogonal to $k'$, as is ${\cal P}$. Furthermore, 
like ${\cal P}$, it can be expressed in terms of the three  four-vectors $N_{L,T,TT}$,
\bea
{\cal P}_{A_Z}^\mu &=&{\cal P}_{A_Z; L}
\, N_L^\mu +{\cal P}_{A_Z; T}\, N_T^\mu +{\cal P}_{A_Z; TT}\, 
N_{TT}^\mu.
\eea
With these definitions, the cross section can be written as
(see Appendix~\ref{app:details} for the derivation and additional details)
\bea
\frac{d^2\sigma_{A_Z}}{d\cos\theta_\pi\,dE_\pi}&=&
\frac{{\cal B}(\tau\to\pi\nu_\tau) m^2_\tau G_F^2|V_{cd}|^2}
{2\pi^2|\vec k\,|(m^2_\tau-m^2_\pi)}
\int_{E^{\rm inf}_\tau-|\vec k\,|}^{E^{\rm sup}_\tau-|\vec k\,|}dq^{0}
\int_{q^2_{-}}^{q^2_{+}}
\frac{dq^2\, F^{A_Z}_{\nu_\tau \Lambda_c}(q^0,q^2)}{
\sqrt{(q^2-q^2_{-})
(q^2_{+}-q^2)}}\no
&\times&
\left[1+\frac{2m_\tau}{m^2_\tau-m^2_\pi}
\Big(\frac{{\cal P}_{A_Z;L}}{m_\tau}(E_\pi |\vec k\,'|-|\vec
p_\pi|k^{\prime0}\cos\theta_{\tau\pi})+
{\cal P}_{A_Z;T}\frac{|\vec p_\pi|}{\sin\theta_\tau}(\cos\theta_\pi-\cos\theta_\tau
\cos\theta_{\tau\pi})\Big)\right]\label{eq:sigmaFinal}
\eea
with $\cos\theta_{\tau\pi}=(2k^{\prime0}E_\pi-m^2_\tau-m^2_\pi)/(2|\vec k\,'|\,|\vec p_\pi|)$ the 
cosine of the angle made by the tau and pion three-momenta, $q^0=k^{\prime0}-|\vec k\,|$ and 
$q^2_{\mp}=m^2_\tau-2|\vec k\,|k^{\prime0}+ 2|\vec k\,|\,|\vec k\,'|\cos(\theta_{\tau\pi}\pm \theta_\pi)$. 
The remaining integration limits, including those for the observed charged pion, namely $E_\pi$ and 
$\cos\theta_\pi$, are discussed in Appendix~\ref{app:limits_integration}.  The nuclear spin-averaged
squared amplitude, $F^{A_Z}_{\nu_\tau \Lambda_c}(q)$, and the polarization-vector projections are
given in Eqs.~\eqref{eq:nucl-ampl}--\eqref{eq:p-projections}. They are expressed in terms of two
ingredients: a set of factors   that depend on both $q^0$ and $|\vec q\,|$ and encode nuclear
effects, being fully determined by the neutron-hole SF, and a set of functions of the variable 
$\omega$, which describes the SM and NP dynamics of the $\nu_\tau n \to \tau^-\Lambda_c^+$ 
transition. As discussed below Eq.~\eqref{eq:polNucl}, for the initial neutron and the final charmed
hyperon embedded in the nuclear medium, we have assumed that the difference between their 
virtualities approximately vanishes, i.e., $(p^2-M_n^2)-(p^{\prime\,2}-M_{\Lambda_c}^2)\approx 0$.

Equation~\eqref{eq:sigmaFinal} can be readily generalized in two directions. First, although we 
explicitly consider only $J^P=1/2^+$ final baryons ($\Lambda_c$ here and $\Lambda$ at the end of 
Sec.~\ref{sec:CP-asy}), final states with arbitrary spin and parity can be treated by evaluating the
corresponding structure functions in terms of the appropriate form 
factors~\cite{Penalva:2019rgt, Penalva:2020xup, Penalva:2021wye, Du:2022ipt}. Second, other $\tau$ 
decay modes can be included. The cross section for $\nu_\tau A_Z\to \Lambda_c X \tau^- (\rho^- 
\nu_\tau)$ follows from the replacements $m_\pi \to m_\rho$ and ${\cal B}(\tau \to \pi \nu_\tau) \to 
{\cal B}(\tau \to \rho \nu_\tau)$, together with a rescaling of the polarization terms in 
Eq.~\eqref{eq:sigmaFinal} by the $\rho$-meson analyzing power, $(m_\tau^2 - 2 m_\rho^2)/(m_\tau^2 + 
2 m_\rho^2) < 1$~\cite{Penalva:2021wye}, while Ref.~\cite{Penalva:2021wye} also provides the 
ingredients needed for the leptonic decays $\tau^-\to\mu^-\bar\nu_\mu\nu_\tau$ and 
$\tau^-\to e^-\bar\nu_e\nu_\tau$.

As can be seen from Eq.~\eqref{eq:sigmaFinal}, integration over the azimuthal angle of the outgoing
pion eliminates the contribution of the transverse--transverse ($TT$) component of the tau 
polarization vector, as in the free-neutron case studied in Ref.~\cite{Hernandez:2026svn}. Since this 
is the only polarization component sensitive to CP-violating NP effects (see 
Eq.~\eqref{eq:p-projections}), the distributions of Eq.~\eqref{eq:sigmaFinal} are blind to CP 
violation. Accessing it requires retaining information on the pion azimuthal angle, as done with 
the asymmetry introduced in Sec.~\ref{sec:CP-asy}.



\section{Azimuthal pion-angle asymmetry and CP-odd NP contributions}
\label{sec:CP-asy}

\begin{figure}[t]
\begin{center}
\rotatebox{270}{\includegraphics[height=.6\textwidth]{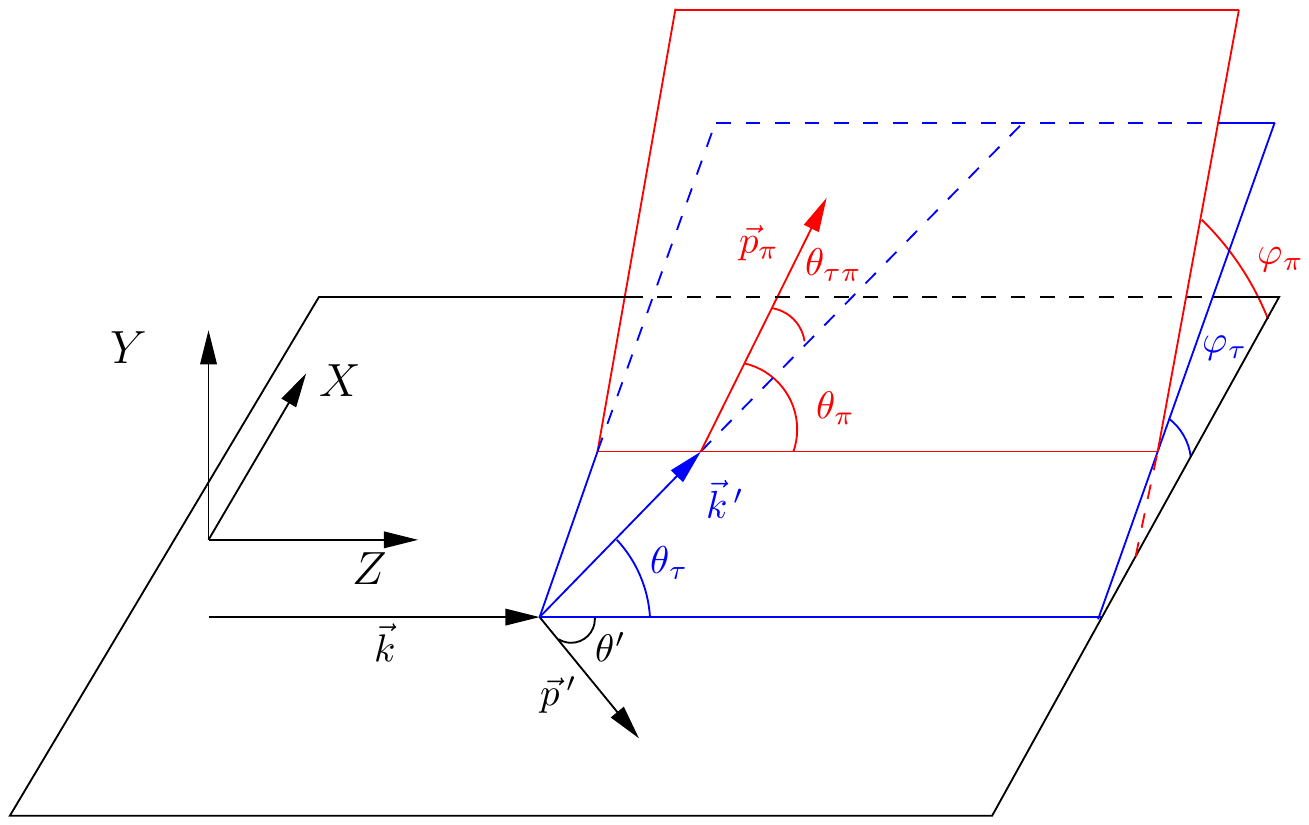}}
\caption{Kinematic configuration used to define the asymmetry in the pion azimuthal angle $\varphi_\pi$. Note that the $\tau$ momentum does not lie in the $OXZ$ plane and, hence, the unit vector $\vec{n}_{TT}=(\vec k\times\vec k\,')/|\vec k\times\vec k\,'|$ does not point along the $Y^+$ axis.}
\label{fig:plano}
\end{center}
\end{figure}
Following Ref.~\cite{Hernandez:2026svn}, we define the asymmetry event by event in a coordinate 
system with the positive $Z$ axis along the incoming neutrino momentum $\vec k$ and the positive 
$Y$ axis along $\vec p\,'\times\vec k$, where $\vec p\,'$ is the outgoing $\Lambda_c^+$ momentum 
(see Fig.~\ref{fig:plano}). The $OXZ$ plane is thus the plane spanned by the incoming neutrino and
the outgoing baryon, and the asymmetry is given by the difference between the cross sections for 
pions emitted above and below it,
\bea
\Delta\sigma_{A_Z}=\sigma_{A_Z}(0<\varphi_\pi<\pi)-\sigma_{A_Z}(\pi<\varphi_\pi<2\pi),\label{eq:asyDef}
\eea
where $\varphi_\pi$ is the pion azimuthal angle in this frame.

For production off a free nucleon, $\vec p\,'=\vec k-\vec k\,'$, so that $\vec p\,'\times\vec k=
\vec k\times\vec k\,'$ and the $OXZ$ plane coincides with the lepton scattering plane. Moreover, 
measuring the final-state baryon momentum at fixed neutrino energy fully determines the $\tau$ 
momentum, so that the axes defining ${\cal P}_L$ and the transverse polarization components are 
known event by event. In Ref.~\cite{Hernandez:2026svn} we showed that, in this case, a nonzero 
$\Delta\sigma$ can only be generated by CP-violating NP. For a nuclear target, in contrast, the 
initial nucleon momentum is unobserved and Fermi motion allows the intermediate $\tau$ lepton to 
emerge outside the $OXZ$ plane, as illustrated in Fig.~\ref{fig:plano}. Since the definition above 
involves only $\vec k$ and $\vec p\,'$, it applies unchanged, with the coordinate system 
reconstructed independently for each event. As shown below, this remains true for nuclear targets within the approximations adopted here, 
namely the impulse approximation for the production process and the neglect of nuclear rescattering
effects on the final hyperon momentum. These approximations are expected to be sufficiently accurate for events in which the 
produced hyperon is highly energetic and no hadronic activity other than the pion is observed.

The final expression is given in Eq.~\eqref{eq:asyfinal}  and is proportional to $\widetilde{\cal N}_{{\cal H}3}(\omega,k\cdot p)$, defined in Eq.~\eqref{eq:nh3}. This quantity is nonzero only in the presence of CP-violating NP beyond the Standard Model~\cite{Penalva:2021gef}, i.e., when the Wilson coefficients are complex and have nonvanishing relative phases with respect to the SM contribution. Consequently, a nonvanishing asymmetry can arise only from such NP.

In contrast to the nucleon-level case studied in Ref.~\cite{Hernandez:2026svn}, where the asymmetry is driven entirely by the ${\cal P}_{TT}$ component of the tau polarization vector, in the present nuclear case all three components, ${\cal P}_{L}$, ${\cal P}_{T}$, and ${\cal P}_{TT}$, contribute. This is because the nonzero momentum of the struck nucleon implies that the reference plane no longer coincides with the lepton plane. Nevertheless, the ${\cal P}_{TT}$ contribution remains dominant, whereas the ${\cal P}_{L}$ and ${\cal P}_{T}$ contributions vanish in the limit in which the characteristic momentum scale of the bound-nucleon momentum distribution is negligible compared with the momentum transfer $\vec q$ from the leptons to the nucleus. See Appendix~\ref{app:asy} for details.

\section{ $\bar\nu_\tau A_Z\to \tau^+(\pi^+\bar\nu_\tau)\Lambda X$ reaction}
\label{sec:anti}
As discussed in Ref.~\cite{Hernandez:2026svn}, all the expressions  for
 the $\nu_\tau n\to \Lambda_c \tau^-$ reaction  can be used for $\bar \nu_\tau p\to \Lambda \tau^+$ 
 with the obvious changes in the CKM matrix element,  masses, form factors and Wilson
coefficients. Besides, for the case where the $\tau^+$ decays into $\pi^+\bar\nu_\tau$, 
one has to implement the extra change
\bea
\Big(1+\frac{2m_\tau}{m^2_\tau-m^2_\pi}\,p_\pi\cdot{\cal P}\Big)\to
\Big(1-\frac{2m_\tau}{m^2_\tau-m^2_\pi}\,p_\pi\cdot{\cal P}\Big)
\label{eq:change}
\eea
with ${\cal P}$ evaluated  as in the case of the $\nu_\tau n\to \Lambda_c \tau^-(\pi^-\nu_\tau)$ 
reaction but now using the $\Lambda$ and proton masses, the corresponding Wilson coefficients  and 
the appropriate form factors for the $p\to \Lambda$ transition. All this applies directly to our 
calculation here where we just sum over neutrons/protons in the initial nucleus.

 \section{Nuclear effects: the local Fermi gas}
\label{sec:LFG}
So far, the hole SF has been left unspecified, and all the expressions derived in the previous
sections are therefore valid for an arbitrary choice of the SF. For the numerical results presented
below, we describe the nuclear system within the LFG approximation, in which the neutron-hole SF in
symmetric nuclear matter takes the form~\cite{Sobczyk:2019uej}
\bea
S_{h}^{\rm LFG}(E,|\vec p\,|)&=&2\int d^3r\,H\left[k_F(r)-|\vec p\,|\right]\delta\left(E-
\sqrt{M^2_n+|\vec p\,|^2}\right)\label{eq:ShLFG}
\eea
with $k_F(r)=(3\pi^2\rho(r)/2)^{1/3}$ the local Fermi momentum and $\rho(r)$ the local
nucleon density distribution (protons plus neutrons). In an isospin non-symmetric environment, one
can replace $k_F\to k_F^{(n)}=(3\pi^2\rho^{(n)}(r))^{1/3}$, involving only the neutron distribution.

Strictly speaking, within the LFG approximation, nucleons move in an attractive central potential,
$-k_F^2(r)/(2M)$, where $M$ is an average nucleon mass. This potential is not explicitly accounted
for in the energy-conservation delta function, $\delta(E-E'-q^0)$, in Eq.~\eqref{eq:sigaz}. This is
nevertheless a sufficiently accurate approximation provided that the final baryon experiences the
same attractive potential, so that the corresponding potential-energy contributions cancel. We are
thus implicitly assuming that the produced $\Lambda_c^+$ hyperon experiences the same local binding
potential as the struck neutron (and, likewise, the $\Lambda$ hyperon as the struck proton). This is
why the on-shell baryon energies used in Eq.~\eqref{eq:sigmaAZ1} do not amount to neglecting the
effects of the nuclear medium on the produced hyperon altogether.\footnote{For the $d\to u$
transition producing a proton rather than a hyperon, Pauli blocking of the final proton should be
taken into account, at a minimum. More generally, the delta function $\delta[E'-(E-q^0)]$ in
Eqs.~\eqref{eq:am}-\eqref{eq:c0d0} should then be replaced by a particle SF~\cite{Nieves:2017lij},
which accounts for the dynamics of the nucleon embedded in the nuclear medium. In that case, all
integrals of Eqs.~\eqref{eq:am}-\eqref{eq:c0d0} are related to the imaginary part of the Lindhard
function (see Appendix B of Ref.~\cite{Nieves:2004wx}).}

Despite its simplicity, the LFG model provides a remarkably realistic description of nuclear momentum
and energy distributions. For $^{12}$C, the two-dimensional energy-momentum distribution predicted
by the LFG model, shown in Fig.~3 of Ref.~\cite{Bourguille:2020bvw}, closely resembles those obtained
using more sophisticated SF approaches~\cite{Benhar:1989aw,Benhar:1994hw,Benhar:2005dj}, and
represents a substantial improvement over the global FG model, which assumes a spatially uniform
Fermi momentum rather than accounting for the local nuclear density. Likewise, the comparison in
Fig.~4 of Ref.~\cite{Alvarez-Ruso:2014bla} of the nucleon momentum distributions in ${^{12}}$C
predicted by the LFG and global FG models with those obtained from the realistic convolution model
of Ref.~\cite{CiofidegliAtti:1995qe} clearly shows that the LFG model reproduces the overall
momentum distribution considerably more accurately than the global FG approximation.

More directly relevant to the present work, Ref.~\cite{Hernandez:2025snr} investigated the inclusive
reactions $\nu_\tau A_Z \to \tau^- \Lambda_c^+ X$ and $\bar\nu_\tau A_Z \to \tau^+ \Lambda X$, from
which the processes studied here originate, using also the hole SFs of Ref.~\cite{Sobczyk:2023mey}.
These are based on realistic nuclear many-body wave functions obtained within coupled-cluster
theory~\cite{Hagen:2013nca} and a nuclear Hamiltonian derived from chiral effective field theory.
The resulting differential and total cross sections are in remarkably good agreement with those
obtained using the LFG SFs, provided that the produced hyperon is assumed to have an average binding
energy similar to that of the nucleon hole, which supports the assumption made above.

For the particular hole SF of Eq.~\eqref{eq:ShLFG}, the factors involving integrations over the SF in
Eq.~\eqref{eq:sigmaFinal} can be obtained analytically, as detailed in
Eqs.~\eqref{eq:aLFGm}--\eqref{eq:dLFGm} of Appendix~\ref{app:LFG}, where some of the integration
limits are also determined more precisely. The corresponding simplifications for the CP asymmetry
are discussed in Appendix~\ref{app:asyLFG}, and the final expression for $\Delta\sigma_{A_Z}$ in
this approximation is given in Eq.~\eqref{eq:assyLFG}.

%

\section{Results}
\label{sec:results}
Equation~\eqref{eq:sigmaFinal}, together with its extension to antineutrino-induced processes,
holds for the complete set of dimension-six four-fermion operators. In this work, however, we
mostly restrict ourselves to SM predictions, since the current uncertainties in the hadronic form
factors and nuclear effects are expected to mask any NP contributions that are small compared to
the SM ones. NP effects are considered in two places: in one illustrative case for the pion energy
distribution, and for the CP asymmetry of Sec.~\ref{sec:CP-asy}, both discussed at the end of this
section.

All calculations are performed for a ${}^{16}$O target within the LFG model introduced in
Sec.~\ref{sec:LFG}. For $\Lambda_c^+$ production, we employ the lattice-QCD form factors of
Ref.~\cite{Meinel:2017ggx}, obtained for the weak decay $\Lambda_c\to n$. For $\Lambda$ production,
we use two different sets, the lattice-QCD~\cite{Bacchio:2025auj} and
phenomenological~\cite{Mintz:2004hm} form factors extracted from the $\Lambda\to p$ weak decay;
unless stated otherwise, the lattice set is used. In all cases, these form factors are only known in the narrow range of small timelike $q^2$ accessible in those decays, so that their application to the present process requires an analytic continuation to the large negative $q^2$ region, for which there is at present no reliable determination.  The associated uncertainties, illustrated for the free-nucleon cross section in the left panel of Fig.~1 of Ref.~\cite{Hernandez:2025snr}, are sizable and are not shown here, so that the predicted cross sections should be regarded as order-of-magnitude estimates.

\begin{figure}[htb]
\begin{center}
\rotatebox{270}{\includegraphics[height=.425\textwidth]{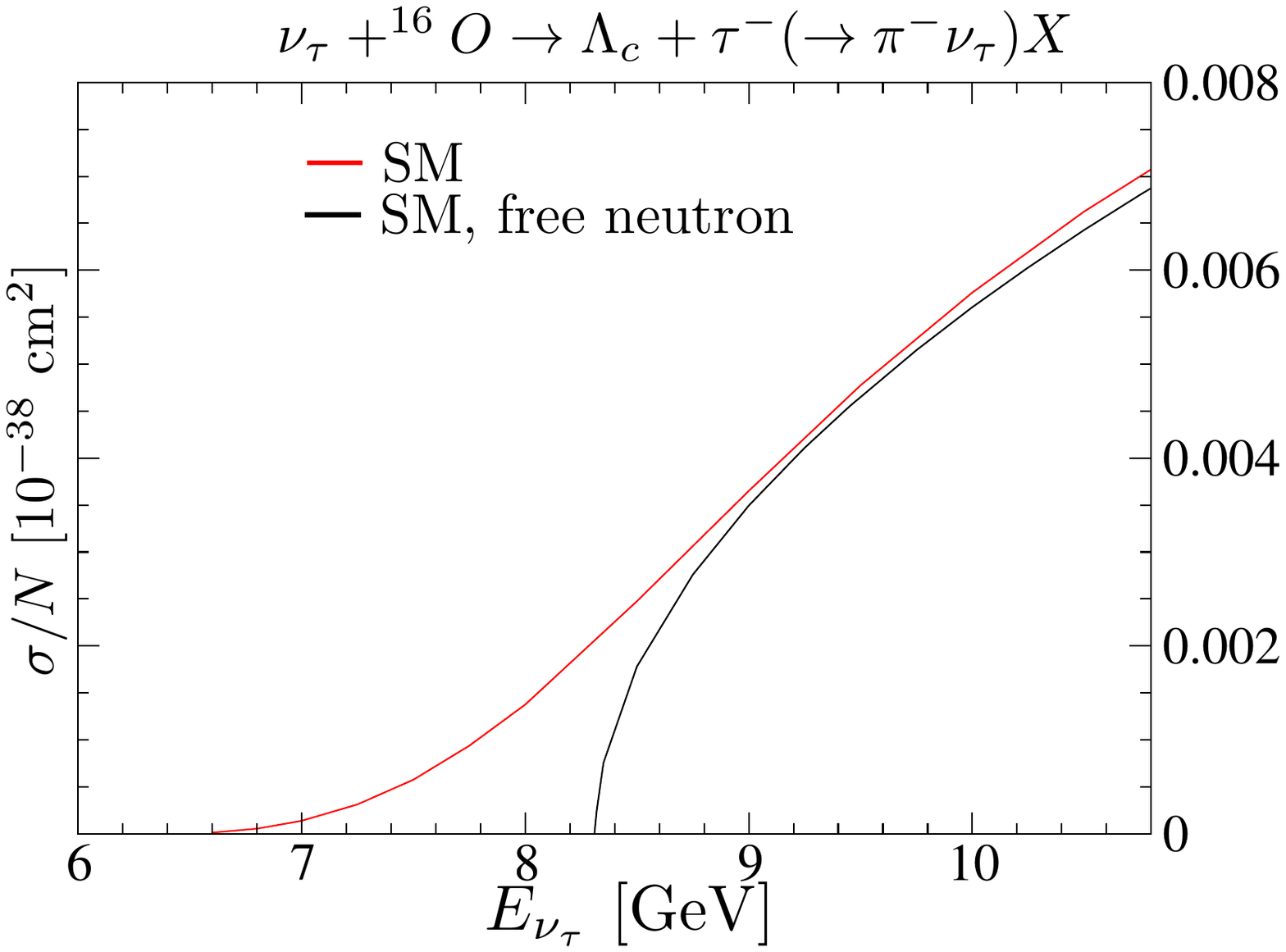}}\ 
\rotatebox{270}{\includegraphics[height=.425\textwidth]{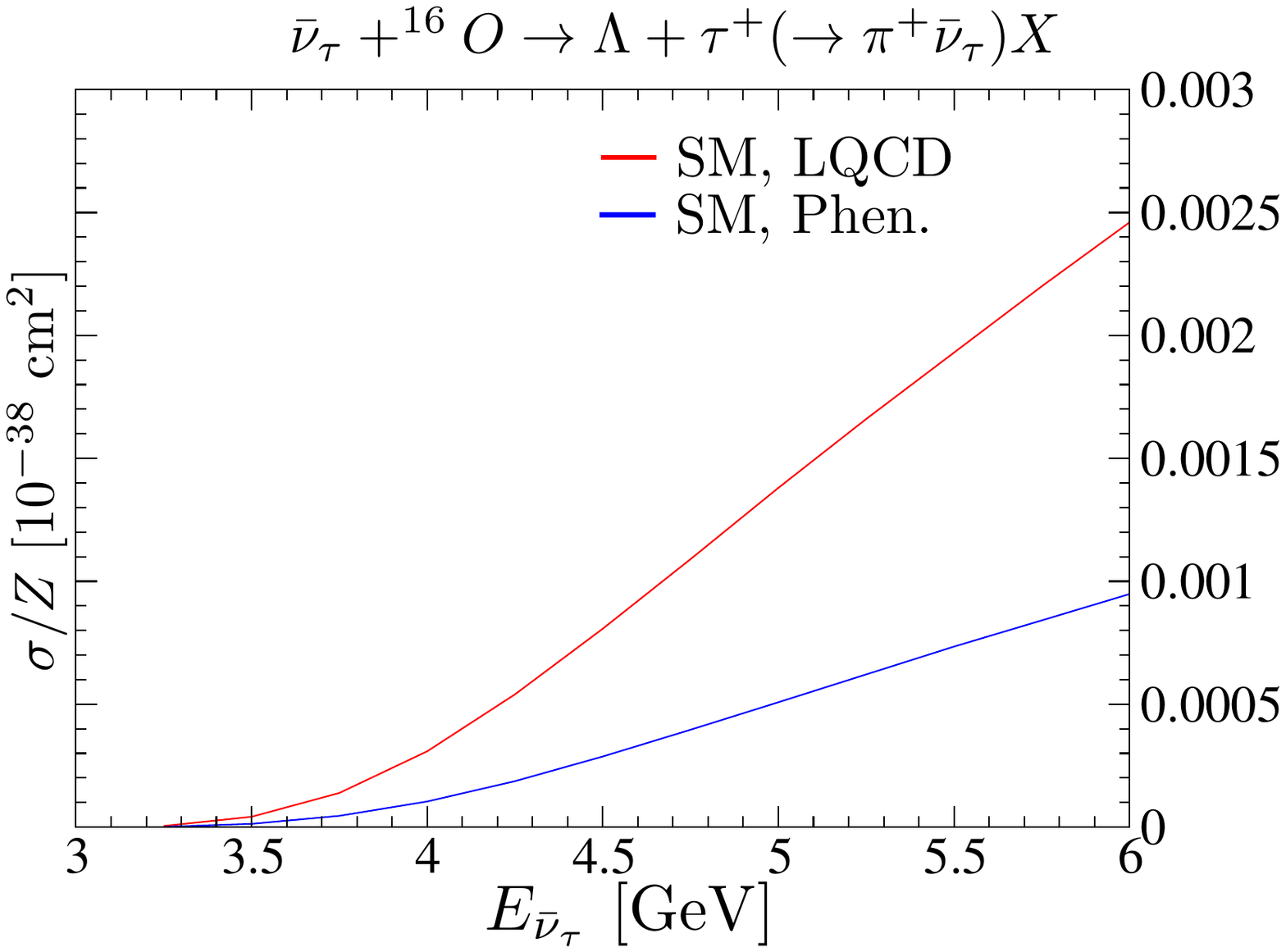}}
\caption{Total cross sections for $\nu_\tau\,{}^{16}\mathrm{O}\to\Lambda_c^+\tau^-(\to\pi^-\nu_\tau)X$ (left panel) and $\bar\nu_\tau\,{}^{16}\mathrm{O}\to\Lambda\tau^+(\to\pi^+\bar\nu_\tau)X$ (right panel), per participant nucleon (neutron or proton, respectively), as functions of the incoming (anti)neutrino energy. In the left panel, the free-neutron cross section is also shown for comparison; in the right panel, results are given for both the lattice-QCD~\cite{Bacchio:2025auj} and the phenomenological~\cite{Mintz:2004hm} form factors.}
\label{fig:total}
\end{center}
\end{figure}

We begin by presenting in Fig.~\ref{fig:total} the total cross section per participant nucleon for
the two reactions $\nu_\tau\,{}^{16}\mathrm{O}\to \Lambda_c\tau^-(\pi^-\nu_\tau)X$ and $\bar\nu_\tau\,{}^{16}\mathrm{O}\to
\Lambda\tau^+(\pi^+\bar\nu_\tau)X$. As follows from Eq.~\eqref{eq:totcs}, these correspond to the
total QE cross section for $\tau$--hyperon production multiplied by the branching ratio for the decay $\tau\to\pi\nu_\tau$.
A robust conclusion is that nuclear effects substantially reduce the reaction thresholds with
respect to the free-nucleon case, where they are $E_{\bar\nu_\tau}\simeq 4.0$~GeV for $\Lambda$
production and $E_{\nu_\tau}\simeq 8.3$~GeV for $\Lambda_c$ production, compared with
$\simeq 3.5$~GeV for $\tau$ production in the QE channel. This can be clearly seen in the left
panel where, for comparison, we also show the cross section for the $\nu_\tau n\to
\Lambda_c^+\tau^-(\to\pi^-\nu_\tau)$ reaction:  the threshold decreases
from about 8.3~GeV to about 6.3~GeV. A similar effect, not shown, is found for the $\Lambda$ 
production case displayed in the right panel, where the threshold decreases from about
4.0~GeV to about 3.0~GeV. In this latter case we see the two form-factor sets that we use lead to 
markedly different cross sections, mainly in normalization rather than in shape.

\begin{figure}[htb]
\begin{center}
\includegraphics[height=.35\textwidth]{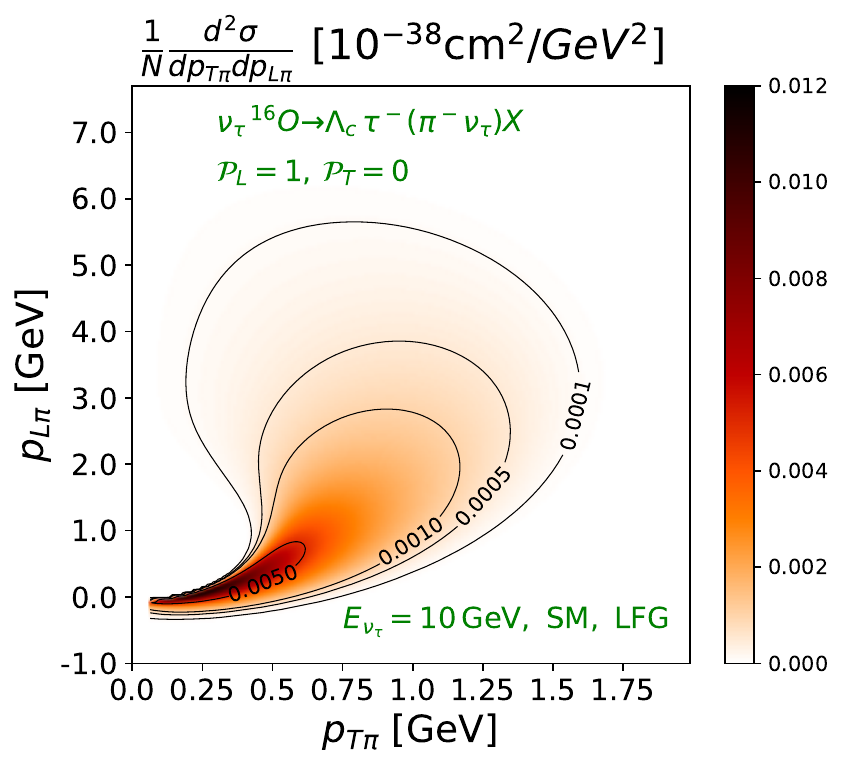}\ 
\includegraphics[height=.35\textwidth]{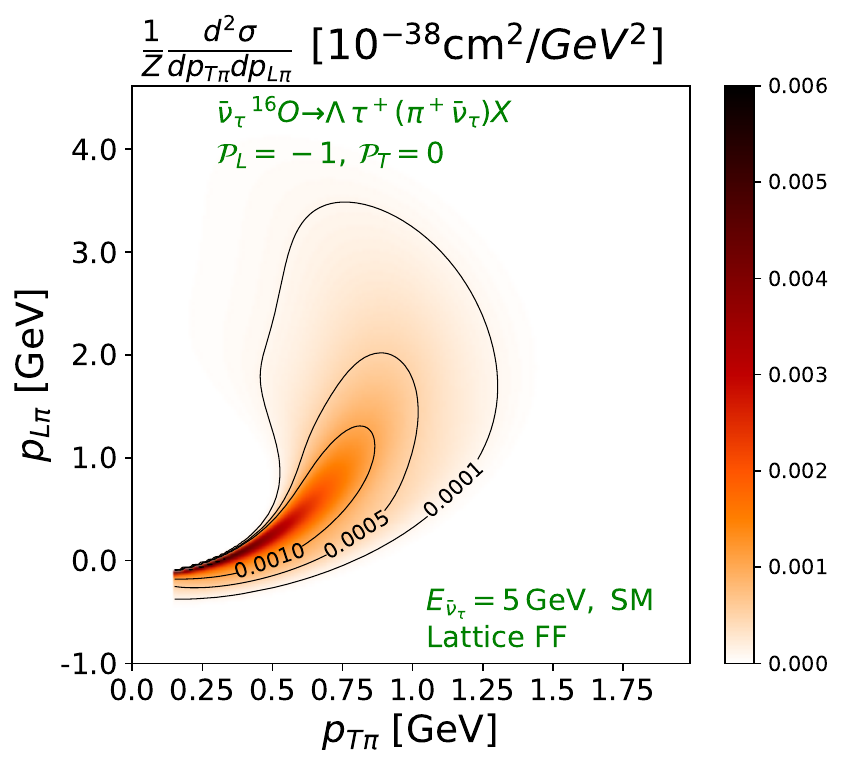}\\
\caption{Double-differential cross section per participant nucleon, $d^2\sigma_{A_Z}/(dp_{T\pi}dp_{L\pi})$
for the $\nu_\tau\,{}^{16}\mathrm{O}\to\Lambda_c^+\tau^-(\pi^-\nu_\tau)X$ (left panel) and 
$\bar\nu_\tau\,{}^{16}\mathrm{O}\to\Lambda\tau^+(\pi^+\bar\nu_\tau)X$ (right panel) reactions. The 
results are obtained within the SM, with nuclear effects described using the LFG model. For $\Lambda$
production, the lattice-QCD form factors from Ref.~\cite{Bacchio:2025auj} are employed.}
\label{fig:ds2dpldpt}
\end{center}
\end{figure}

\begin{figure}[htb]
\begin{center}
\rotatebox{270}{\includegraphics[height=.425\textwidth]{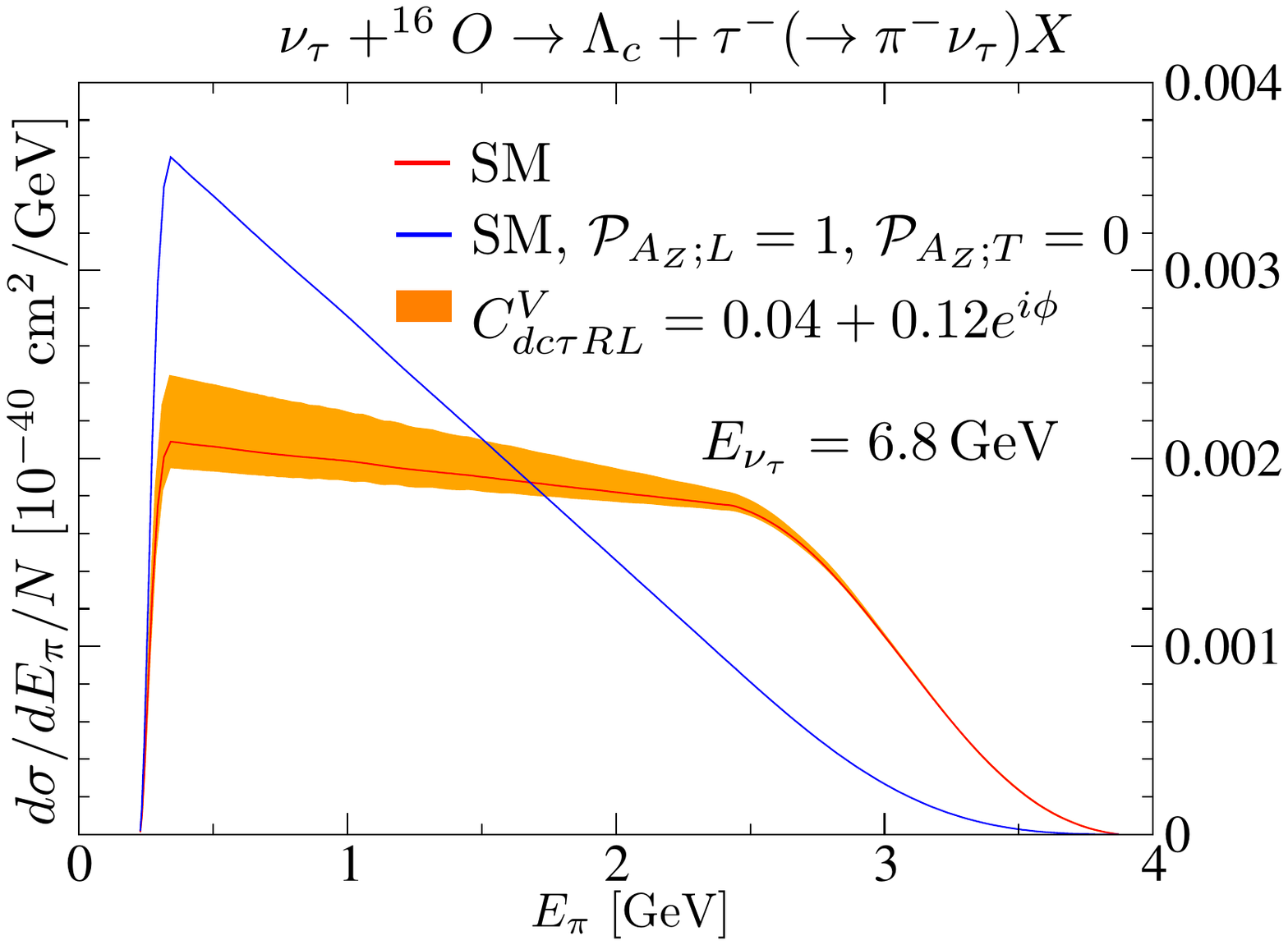}}\ 
\rotatebox{270}{\includegraphics[height=.425\textwidth]{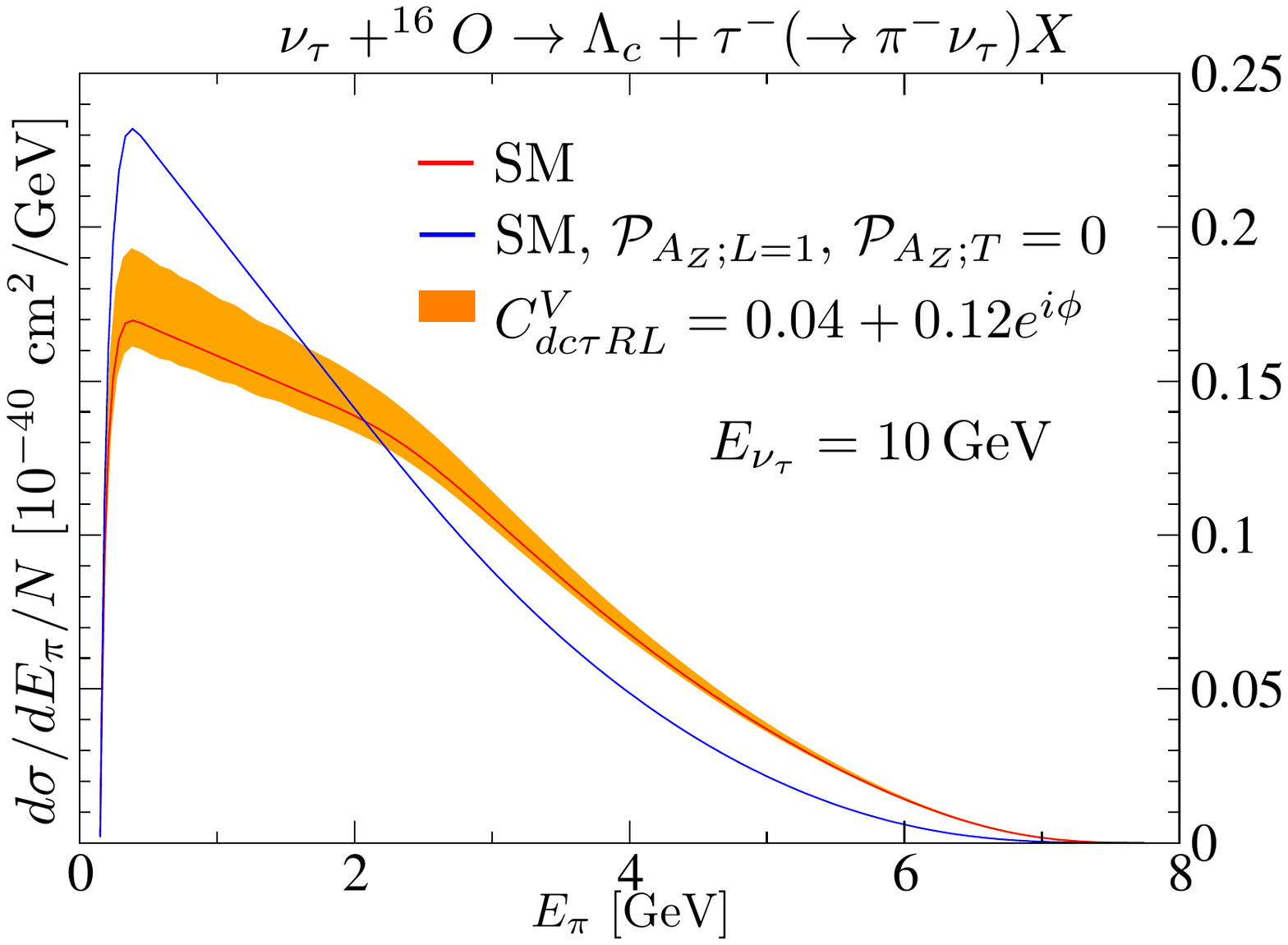}} 
\caption{Differential cross section $d\sigma_{A_Z}/dE_\pi$, per participant neutron, for the reaction 
$\nu_\tau\,{}^{16}\mathrm{O}\to \Lambda_c\tau^-(\pi^-\nu_\tau)X$ at incident neutrino energies of 6.8 GeV 
(left panel) and 10 GeV (right panel). The full calculation (red curve) is compared with the   approximation obtained by setting ${\cal P}_{A_Z;L}=1$ and ${\cal P}_{A_Z;T}=0$ (blue curve) in  Eq.~\eqref{eq:sigmaFinal}. We also show results including NP effects (orange-shaded
band) induced by a nonzero Wilson coefficient $C^V_{dc\tau RL}=0.04 +0.12e^{i\phi}$. In both panels, the upper limit of the band is reached for $\phi=0$, while for the lower limit $\phi=\pi$. } 
\label{fig:dsdepilc}
\end{center}
\end{figure}

\begin{figure}[htb]
\begin{center}
\rotatebox{270}{\includegraphics[height=.425\textwidth]{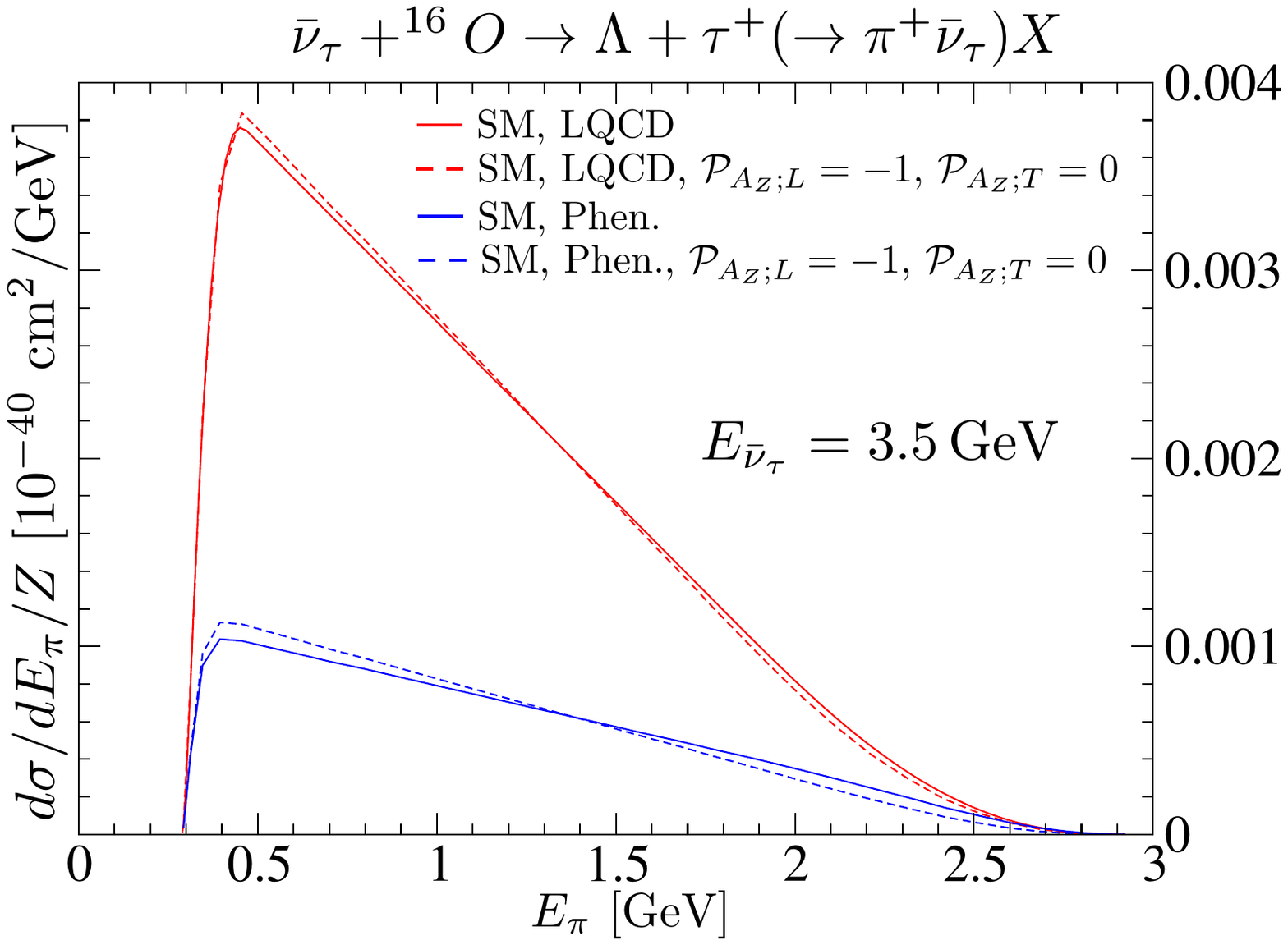}}\ 
\rotatebox{270}{\includegraphics[height=.425\textwidth]{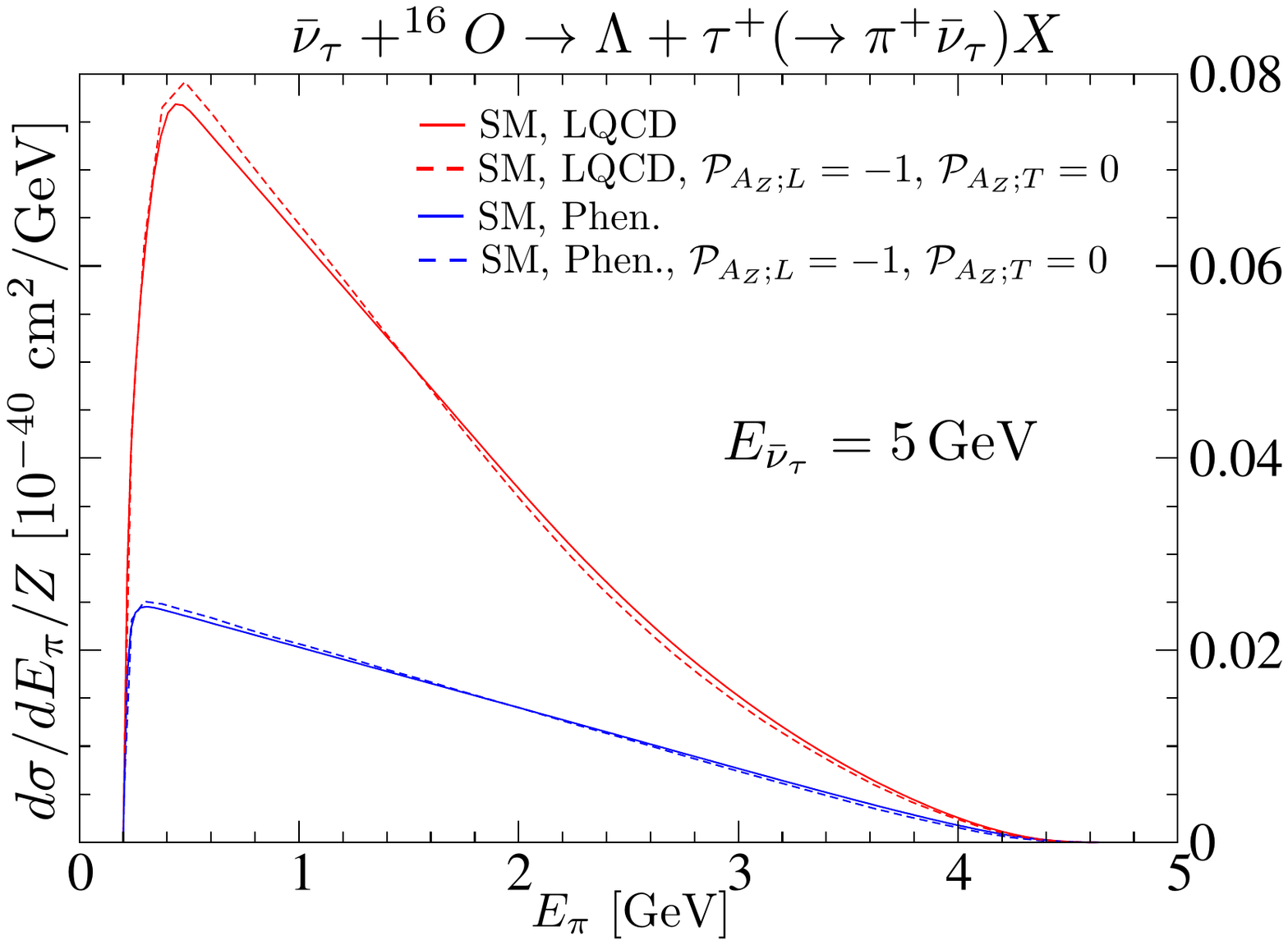}} 
\caption{Differential cross section $d\sigma_{A_Z}/dE_\pi$, per participant proton, for the reaction 
$\bar\nu_\tau\,{}^{16}\mathrm{O}\to \Lambda\tau^+(\pi^+\bar\nu_\tau)X$, obtained with the 
lattice-QCD~\cite{Bacchio:2025auj} and phenomenological~\cite{Mintz:2004hm} form factors, at incident
antineutrino energies of 3.5 GeV (left panel) and 5 GeV (right panel). The full  calculation (solid 
curve) is compared with the approximation obtained by setting ${\cal P}_{A_Z;L}=-1$ and ${\cal P}
_{A_Z;T}=0$ (dashed curve) in Eq.~\eqref{eq:sigmaFinal}. }
\label{fig:dsdepil}
\end{center}
\end{figure}

\begin{figure}[htb]
\begin{center}
\rotatebox{270}{\includegraphics[height=.425\textwidth]{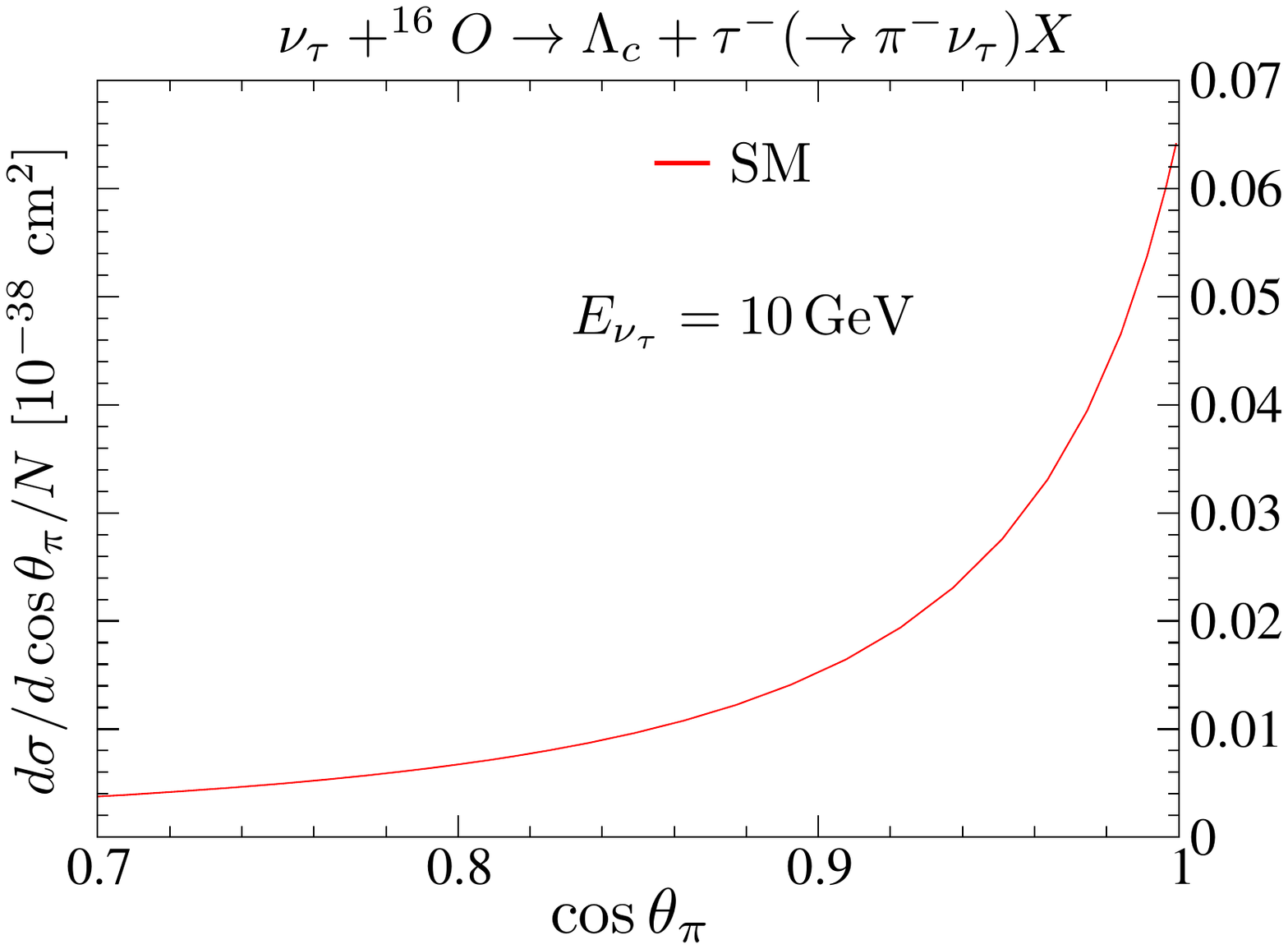}}\
\rotatebox{270}{\includegraphics[height=.425\textwidth]{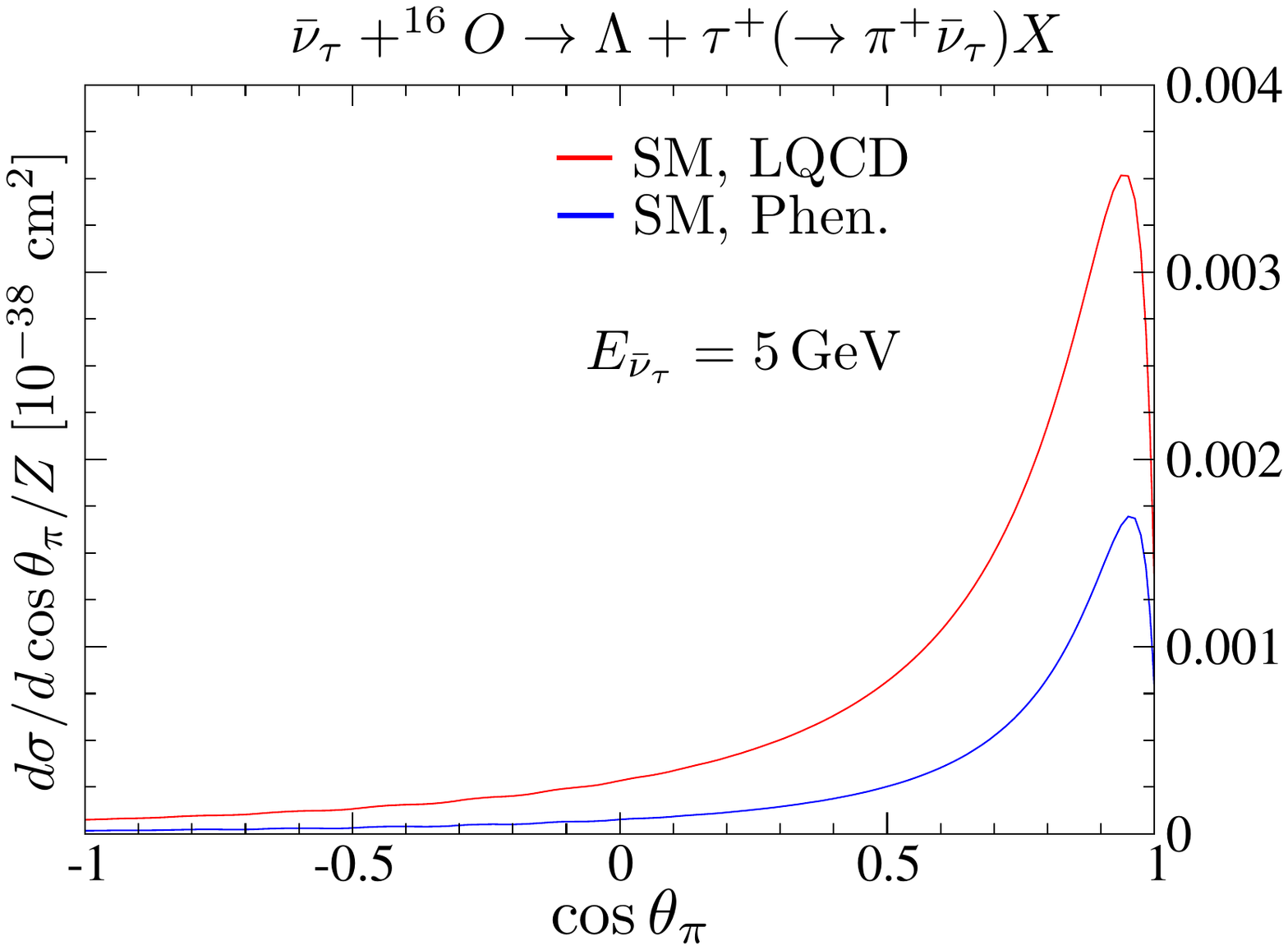}}
\caption{Differential cross section $d\sigma_{A_Z}/d\cos\theta_\pi$, per participant nucleon,  for 
the reactions $\nu_\tau\,{}^{16}\mathrm{O}\to \Lambda_c\tau^-(\pi^-\nu_\tau)X$ (left panel) and 
$\bar\nu_\tau\,{}^{16}\mathrm{O}\to \Lambda\tau^+(\pi^+\bar\nu_\tau)X$ (right panel), evaluated at incident 
neutrino (antineutrino) energies of 10 and 5 GeV, respectively. For the $\Lambda$ production 
channel, the differential cross section is computed using both the lattice-QCD form factors of 
Ref.~\cite{Bacchio:2025auj} and the phenomenological form factors of Ref.~\cite{Mintz:2004hm}. }
\label{fig:dsdctpilcl}
\end{center}
\end{figure}

In Fig.~\ref{fig:ds2dpldpt}, we present the double-differential cross section per participant nucleon,
$d^2\sigma_{A_Z}/dp_{T\pi}dp_{L\pi}$, where $p_{T\pi}$ and $p_{L\pi}$ denote the transverse and
longitudinal pion momenta, defined as
\bea
p_{T\pi}=|\vec p_\pi|\sin\theta_\pi,\qquad
p_{L\pi}=|\vec p_\pi|\cos\theta_\pi.
\eea
These variables are commonly used in experimental analyses, and the corresponding cross section is
related to the energy--angle distribution through~\cite{Hernandez:2022nmp}
\bea
\frac{d^2\sigma_{A_Z}}{dp_{T\pi}dp_{L\pi}}
=\frac{\sin\theta_\pi}{E_\pi}
\frac{d^2\sigma_{A_Z}}{d\cos\theta_\pi dE_\pi}.
\eea
The cross section is seen to be concentrated within a narrow band in the $(p_{T\pi},p_{L\pi})$ plane.
Further insight into this behaviour can be gained from the single-differential distributions
$d\sigma_{A_Z}/dE_\pi$ and $d\sigma_{A_Z}/d\cos\theta_\pi$, shown in Figs.~\ref{fig:dsdepilc}--\ref{fig:dsdctpilcl}.

In Fig.~\ref{fig:dsdepilc}, we show the differential cross section $d\sigma_{A_Z}/dE_\pi$ for the
neutrino reaction $\nu_\tau\,{}^{16}\mathrm{O}\to \Lambda_c^+\tau^-(\pi^-\nu_\tau)X$ at two incident neutrino
energies, 6.8~GeV, just above threshold, and 10~GeV. For comparison, we also display the results
obtained by setting ${\cal P}_{A_Z;L}=1$ and ${\cal P}_{A_Z;T}=0$ in Eq.~\eqref{eq:sigmaFinal},
corresponding to the limiting case of an ultra-relativistic $\tau$. The comparison highlights the
importance of properly accounting for the $\tau$ polarization. In particular, the physical
distribution is less sharply peaked towards low pion energies and, at the lower incident neutrino
energy, develops a broad plateau over which the differential cross section exhibits only a mild
variation.

The situation is different for the $\bar\nu_\tau\,{}^{16}\mathrm{O}\to \Lambda\tau^+(\pi^+\bar\nu_\tau)X$
antineutrino reaction, whose $d\sigma_{A_Z}/dE_\pi$ distribution is shown in Fig.~\ref{fig:dsdepil}
for both form-factor sets. In this case, imposing ${\cal P}_{A_Z;L}=-1$ and ${\cal P}_{A_Z;T}=0$
(appropriate for a $\tau^+$) has only a minor effect on the results. This is somewhat unexpected,
since the produced $\tau^+$ leptons are, on average, less energetic than those in the $\Lambda_c$
production channel. Besides possible form-factor effects, we attribute this behaviour to the sign
change of the vector--axial interference term in the contraction of the hadronic and leptonic
tensors when going from neutrino- to antineutrino-induced reactions, which makes the $\tau^+$
angular distribution more forward peaked.

Finally, in Fig.~\ref{fig:dsdctpilcl}, we present the per-participant-nucleon differential cross
sections $d\sigma_{A_Z}/d\cos\theta_\pi$. The distributions are strongly forward peaked,
especially for $\Lambda_c^+$ production. Both the $d\sigma_{A_Z}/dE_\pi$ and
$d\sigma_{A_Z}/d\cos\theta_\pi$ distributions per participant nucleon closely resemble their
free-nucleon counterparts reported in Ref.~\cite{Hernandez:2026svn}. This similarity is readily
understood: the incoming neutrino traverses the nucleus without undergoing strong interactions,
while the $\tau$ decays outside the nucleus, producing a pion and a neutrino. Thus, nuclear effects
primarily affect the initial hyperon-production process, with little impact on the subsequent pion
distributions resulting from the $\tau$ decay.

We now turn to NP effects. In Fig.~\ref{fig:dsdepilc}, for the $\Lambda_c$ production case, the
orange band shows the results obtained for $C^V_{dc\tau RL}=0.04+0.12e^{i\phi}$. As discussed in
Ref.~\cite{Hernandez:2025snr}, this is the only $d\to c\tau$ Wilson coefficient that is not strongly
constrained to small values by the combined analysis of Drell--Yan data and purely leptonic
$D$-meson decays performed in Ref.~\cite{Fuentes-Martin:2020lea}. Following
Ref.~\cite{Hernandez:2025snr}, we assume that the deviations from the SM remain moderate and take
$|C^V_{dc\tau RL} - 0.04| \simeq 0.12$, ensuring that $|C^V_{dc\tau RL}|={\cal O}(10^{-1})$ while,
at the same time, staying consistent with the constraints derived in
Ref.~\cite{Fuentes-Martin:2020lea}. As can be seen, this Wilson coefficient largely preserves the
overall shape of the distribution while changing its magnitude.

Tables~\ref{tab:asy} and \ref{tab:asylambda} show the relative CP asymmetry $\Delta\sigma_{A_Z}/\sigma_{A_Z}$, defined in Eq.~\eqref{eq:asyDef}, together with the variation of the total cross section induced by the NP. The asymmetry is evaluated with the LFG expression of Eq.~\eqref{eq:assyLFG}, obtained from the general formula in Eq.~\eqref{eq:asyfinal}. As in our previous nucleon-level study~\cite{Hernandez:2026svn}, we adopt
the same CP-violating Wilson coefficients and parameter values, and refer the reader to that work for
their motivation. The nuclear results are broadly consistent with those obtained for free-nucleon
targets. In the $\Lambda_c$ channel, the asymmetry remains below $0.5\%$ for the Wilson coefficients
considered, whereas in the $\Lambda$ channel it reaches the percent level, with a magnitude, and for
$C^S_{su\tau LL}$ even a sign, that depends strongly on the form-factor set. Only its order of
magnitude should therefore be regarded as meaningful.
\begin{table}[h!]
\begin{center}
\begin{tabular}{|c|c|c|}\hline
      $C^S_{dc\tau LL}$      &  $\frac{\Delta\sigma_{A_Z}}{\sigma_{A_Z}}(\%)$ & 
      $\frac{\sigma_{A_Z}-\sigma^{\rm SM}_{A_Z}}{\sigma^{\rm SM}_{A_Z}}(\%)$\\\hline
$0.01i$&$-0.14$& $2.8\times10^{-3}$\\ 
$0.03i$&$-0.41$&$2.5\times10^{-2}$\\\hline 
\end{tabular}
\begin{tabular}{|c|c|c|}\hline
      $C^S_{dc\tau RL}$      &  $\frac{\Delta\sigma_{A_Z}}{\sigma_{A_Z}}(\%)$ & 
      $\frac{\sigma_{A_Z}-\sigma^{\rm SM}_{A_Z}}{\sigma^{\rm SM}_{A_Z}}(\%)$\\\hline
$0.01i$&$-0.078$& $2.8\times10^{-3}$\\ 
$0.03i$&$-0.23$&$2.5\times10^{-2}$\\\hline \end{tabular}
\begin{tabular}{|c|c|c|}\hline
      $C^T_{dc\tau LL}$      &  $\frac{\Delta\sigma_{A_Z}}{\sigma_{A_Z}}(\%)$ & 
      $\frac{\sigma_{A_Z}-\sigma^{\rm SM}_{A_Z}}{\sigma^{\rm SM}_{A_Z}}(\%)$\\\hline
$0.001i$&$-3.7\times10^{-3}$& $6.9\times10^{-4}$\\ 
$0.007i$&$-2.6\times10^{-2}$&$3.3\times10^{-2}$\\\hline \end{tabular}
\end{center}
\caption{CP asymmetry $\Delta\sigma_{A_Z}$ and variation of the total cross section, $\sigma_{A_Z}-\sigma^{\rm SM}_{A_Z}$, for the quasielastic $\nu_\tau\,{}^{16}\mathrm{O}\to\Lambda_c^+\,\tau^-(\pi^-\nu_\tau)X$ reaction, evaluated at $E_\nu=10\,$GeV for a fully left-handed $\nu_\tau$ beam. The central value of the SM total cross section per neutron is $\sigma^{\rm SM}_{A_Z}/N=0.00576\times10^{-38}\,{\rm cm}^2$. The NP effects are generated by assuming that only one of the Wilson coefficients $C^S_{dc\tau LL}$, $C^S_{dc\tau RL}$ or $C^T_{dc\tau LL}$ is nonzero and purely imaginary.}
\label{tab:asy}
\end{table} 

\begin{table}[tbh]
\begin{center}
\begin{tabular}{|c|c|c|c|c|}\hline
&\multicolumn{2}{c|}{FFs from Ref.~\cite{Bacchio:2025auj}}&\multicolumn{2}{c|}{FFs from Ref.~\cite{Mintz:2004hm}}\\ \hline
$C^S_{su\tau LL}$      &  $\frac{\Delta\sigma_{A_Z}}{\sigma_{A_Z}}(\%)$ & 
      $\frac{\sigma_{A_Z}-\sigma^{\rm SM}_{A_Z}}{\sigma^{\rm SM}_{A_Z}}(\%)$&  $\frac{\Delta\sigma_{A_Z}}{\sigma_{A_Z}}(\%)$ & 
      $\frac{\sigma_{A_Z}-\sigma^{\rm SM}_{A_Z}}{\sigma^{\rm SM}_{A_Z}}(\%)$\\\hline
$0.01i$&$-0.25$& 0.56&$0.11$&0.009\\ 
$0.1i$&$-1.63$&56&$1.10$&0.90\\\hline 
\end{tabular}
\begin{tabular}{|c|c|c|c|c|}\hline
&\multicolumn{2}{c|}{FFs from Ref.~\cite{Bacchio:2025auj}}&\multicolumn{2}{c|}{FFs from Ref.~\cite{Mintz:2004hm}}\\\hline
$C^S_{su\tau RL}$      &  $\frac{\Delta\sigma_{A_Z}}{\sigma_{A_Z}}(\%)$ & 
      $\frac{\sigma_{A_Z}-\sigma^{\rm SM}_{A_Z}}{\sigma^{\rm SM}_{A_Z}}(\%)$&  $\frac{\Delta\sigma_{A_Z}}{\sigma_{A_Z}}(\%)$ & 
      $\frac{\sigma_{A_Z}-\sigma^{\rm SM}_{A_Z}}{\sigma^{\rm SM}_{A_Z}}(\%)$\\\hline
$0.01i$&$0.43$& 0.56&$0.12$&0.009\\ 
$0.1i$&$2.74$&56&$1.23$&0.90\\\hline 
\end{tabular}
\end{center}
\caption{CP asymmetry $\Delta\sigma_{A_Z}$ and variation of the total cross section, $\sigma_{A_Z}-
\sigma^{\rm SM}_{A_Z}$, for the $\bar\nu_\tau\,{}^{16}\mathrm{O}\to\Lambda\,\tau^+(\pi^+\bar\nu_\tau)X$ 
reaction, evaluated at $E_{\bar\nu}=5\,$GeV for a fully right-handed $\bar\nu_\tau$ beam. The NP 
effects are generated by assuming that either $C^S_{su\tau LL}$ or $C^S_{su\tau RL}$ is nonzero and 
purely imaginary. Two sets of hadronic form factors are considered: the lattice QCD form factors of 
Ref.~\cite{Bacchio:2025auj} and the phenomenological form factors of Ref.~\cite{Mintz:2004hm}. They 
yield SM total cross sections per proton of $\sigma^{\rm SM}_{A_Z}/Z=0.00138\times10^{-38}\,{\rm cm}
^2$ and $\sigma^{\rm SM}_{A_Z}/Z=0.00051\times10^{-38}\,{\rm cm}^2$, respectively. See 
Ref.~\cite{Hernandez:2026svn} for further details.}
\label{tab:asylambda}
\end{table} 

The tables also show how the asymmetry scales with the NP coefficient. For purely
imaginary Wilson coefficients, $\Delta\sigma_{A_Z}$ grows linearly with $|C|$, as expected
from its interference with the SM amplitude, whereas the change in the total cross section
is quadratic. For $C^S_{dc\tau LL}$ in Table~\ref{tab:asy}, going from $0.01i$ to $0.03i$ increases the
asymmetry by a factor $\simeq 3$ and the cross-section variation by a factor $\simeq 9$.
For small Wilson coefficients the asymmetry is a much more sensitive
probe than the rate: for $C^S_{dc\tau LL}=0.01i$ the asymmetry is $-0.14\%$, while the
total cross section changes by only $2.8\times10^{-3}\,\%$. Since the asymmetry is
normalized to the NP-modified cross section, it grows more slowly than linearly once the
NP contribution to the rate becomes large, as in Table~\ref{tab:asylambda} for the lattice form factors
and $C^S_{dc\tau LL}=0.1i$, where the cross section increases by 56\%.  This latter result, already 
found in  \cite{Hernandez:2026svn}, is unexpected given the small values considered for the NP
Wilson coefficients and points to issues with the extrapolation of some form factors to large 
negative $q^2$ values, as discussed in some more detail in that reference. This could explain why 
the pattern observed in Table I is not well satisfied in Table II when LQCD form factors are used.

\section{Conclusions}
\label{sec:conclusions}

We have developed, within the impulse approximation, a formalism for the sequential reactions
$\nu_\tau\,{}^A Z \to \tau^-(\to\pi^-\nu_\tau)\,Y^+ X$ and $\bar\nu_\tau\,{}^A Z \to \tau^+(\to\pi^+
\bar\nu_\tau)\,Y^0 X$, as required to constrain NP in $\tau$-neutrino charged-current interactions
with the nuclear targets employed in neutrino experiments. The resulting expressions accommodate an arbitrary neutrino spin-density matrix and the complete set of dimension-six semileptonic operators, including those involving right-handed neutrinos, with the nuclear structure encoded in the hole SF. In the numerical applications, however, we assume SM-like fully polarized beams, for which operators involving right-handed neutrinos do not contribute at the primary vertex. Since the $\tau$ decays before it can be tracked, its polarization must be inferred from the decay pion. The double-differential cross section for the pion (Eq.~\eqref{eq:sigmaFinal}) consistently incorporates the longitudinal and transverse components of this polarization, averaged over the nuclear initial state.

We have also constructed an azimuthal pion asymmetry, Eq.~\eqref{eq:asyfinal}, which generalises the
free-nucleon observable of Ref.~\cite{Hernandez:2026svn}. It is defined event by event with respect
to the plane spanned by the incoming neutrino and the outgoing baryon, even though Fermi motion
allows the intermediate $\tau$ to have a momentum component normal to this plane. It vanishes
identically in the SM so that, within the impulse approximation, any nonvanishing asymmetry would
provide a direct signal of CP-violating NP. In contrast to the free-nucleon case, all three
components of the $\tau$ polarization vector contribute, although ${\cal P}_{TT}$ remains dominant.

Our numerical results for $\Lambda$ and $\Lambda_c$ production off ${}^{16}$O, obtained within a LFG
model and focused on SM predictions, show that nuclear effects substantially lower the reaction thresholds, from approximately
$8.3$ to $6.3$~GeV and from $4.0$ to $3.0$~GeV for $\Lambda_c$ and $\Lambda$ production, respectively.
In contrast, the per-participant-nucleon distributions $d\sigma/dE_\pi$ and $d\sigma/d\cos\theta_\pi$ closely resemble their free-nucleon counterparts, and are strongly forward peaked, particularly for $\Lambda_c$ production. This is because the neutrino crosses the nucleus without strong interactions and the $\tau$ decays outside it, so nuclear effects act mainly on the baryon-production vertex. An accurate treatment of the $\tau$ polarization is important for
$\Lambda_c$ production: compared with the ultra-relativistic approximation, the pion spectrum is less
sharply peaked at low energies and develops a broad plateau at the lower incident neutrino energies,
whereas the effect is much smaller for $\Lambda$ production induced by antineutrinos. An illustrative NP scenario for $\Lambda_c^+$ production, $C^V_{dc\tau RL} = 0.04 + 0.12 e^{i\phi}$, largely preserves the shape of the pion energy distribution while changing its magnitude. 
Finally, for
the representative CP-violating NP scenarios of Ref.~\cite{Hernandez:2026svn}, the predicted
asymmetries are small, at most at the percent level for $\Lambda$ production and below $0.5\%$ for
$\Lambda_c$ production, and broadly consistent with the free-nucleon results.

Several limitations affect the quantitative reliability of these predictions. First, the transition
form factors are poorly constrained in the large spacelike-$q^2$ region relevant here: their analytic
continuation introduces sizable uncertainties in both channels, and for $\Lambda$ production the
lattice-QCD and phenomenological form factors yield markedly different normalizations and CP
asymmetries. The predicted cross sections, and the asymmetries in the $\Lambda$ channel, should
therefore be regarded as order-of-magnitude estimates. Second, our calculation only partially accounts for the nuclear dynamics of the outgoing baryon
and relies on an approximate treatment of the off-shell behaviour of the struck nucleon; in
particular, we assume that the produced hyperon experiences the same binding potential as the
initial nucleon. A quantitative assessment of the impact of rescattering effects on the
asymmetry, which is a naive-$T$-odd observable, is left for future work.
The impact of some of these nuclear
effects could be reduced by selecting events with highly energetic outgoing baryons and no
detectable hadronic activity beyond the delayed pion from the $\tau$ decay.

The formalism developed here provides a general framework for investigating NP scenarios and can be
readily combined with more realistic descriptions of nuclear structure, such as the 
coupled-cluster SFs of Ref.~\cite{Sobczyk:2023mey}. It can also be extended to final baryons of arbitrary spin and parity, and
to other $\tau$ decay channels, such as $\tau\to\rho\nu_\tau$, $\tau\to\mu\bar\nu_\mu\nu_\tau$ and
$\tau\to e\bar\nu_e\nu_\tau$. These developments provide a basis for exploiting $\tau$-neutrino data
from DUNE, SHiP and forward experiments at the LHC to probe NP in the $d\to c$ and $u\to s$
transitions, provided the associated hadronic uncertainties can be brought under better control.

\section*{Acknowledgements}
This work is supported by Grants PID2023-147458NB-C21 and CEX2023-001292-S, funded by the European Regional Development Fund (ERDF/EU) and MICIU/AEI (10.13039/501100011033); Grant {\small CIPROM/2023/59}, funded by the Generalitat Valenciana (10.13039/501100003359); the HADRON-2030 project, funded by the European Union's Horizon Europe Research and Innovation Programme under Grant Agreement No. 101292522; and Grant SA091P24, funded by the Junta de Castilla y León under Programme EDU/841/2024.


\appendix

\section{Some details of phase-space integrations}
\label{app:details}

With the definitions of Eqs.~\eqref{eq:nuctenquasie}--\eqref{eq:faz}, the cross section of Eq.~\eqref{eq:sigaz} can be written as
\bea
\sigma_{A_Z}
&=&\frac{{\cal B}(\tau\to\pi\nu_\tau) m^2_\tau G_F^2|V_{cd}|^2}
{8\pi^3|\vec k\,|(m^2_\tau-m^2_\pi)}
\int\frac{d^3k'F^{A_Z}_{\nu_\tau \Lambda_c}(q)}{k^{\prime0}}\int d^3p_\pi
\frac{\delta(k^{\prime0}-E_\pi-|\vec k\,'-\vec p_\pi|)}{ E_\pi\,|\vec k\,'-\vec p_\pi|}
\Big(1+\frac{2m_\tau}{m^2_\tau-m^2_\pi}
\,p_\pi\cdot{\cal P}_{A_Z}\Big).
\eea
For each value of $\vec p_\pi$, let $\tilde R$ denote a rotation about the $Z$ axis such that
$\tilde p_\pi = \tilde R p_\pi =
\left(E_\pi, |\vec p_\pi|\sin\theta_\pi, 0,
|\vec p_\pi|\cos\theta_\pi\right)$
Exploiting rotational invariance and performing the change of integration
variable $\tilde R\vec k\,' \to \vec k\,'$, we obtain 
\bea
\sigma_{A_Z} &
=& \frac{{\cal B}(\tau\to\pi\nu_\tau) m^2_\tau G_F^2|V_{cd}|^2}
{4\pi^2|\vec k\,|(m^2_\tau-m^2_\pi)}
\int d(\cos\theta_\pi) dE_\pi |\vec p_\pi| \int\frac{d^3k'}{ k^{\prime0}\,|\vec
k\,'-\vec {\tilde p}_\pi|}
\delta(k^{\prime0}-E_\pi-|\vec k\,'-\vec {\tilde p}_\pi|)\no
&\times&
F^{A_Z}_{\nu_\tau \Lambda_c}(q)\Big(1+\frac{2m_\tau}{m^2_\tau-m^2_\pi}
\,\tilde p_\pi\cdot{\cal P}_{A_Z}\Big). \label{eq:A2}
\eea
Using the delta function, the integration over the tau-lepton azimuthal
angle, $\varphi_\tau$, can be performed, yielding
\bea
\sigma_{A_Z}
&=&\frac{{\cal B}(\tau\to\pi\nu_\tau) m^2_\tau G_F^2|V_{cd}|^2}
{2\pi^2|\vec k\,|(m^2_\tau-m^2_\pi)}
\int d(\cos\theta_\pi) dE_\pi\int\frac{dk^{\prime0}\,d(\cos\theta_\tau)}{
\sin\theta_\tau\sin\theta_\pi|\sin\varphi^0_\tau|}
H\left[1-|\cos\varphi^0_\tau|\right]F^{A_Z}_{\nu_\tau \Lambda_c}(q)\no
&\times& \left[1+\frac{2m_\tau}{m^2_\tau-m^2_\pi}
\left(\frac{{\cal P}_{A_Z; L}}{m_\tau}\left(E_\pi |\vec k\,'|-|\vec
p_\pi|k^{\prime0}\cos\theta_{\tau\pi}\right)+
\frac{{\cal P}_{A_Z; T}|\vec p_\pi|}{\sin\theta_\tau}\left(\cos\theta_\pi-\cos\theta_\tau
\cos\theta_{\tau\pi}\right)\right)\right], \label{eq:sA}
\eea
%
 where $H[\cdots]$ denotes the Heaviside step function, $\theta_{\tau\pi}$ is the angle between
the $\tau$ and pion three-momenta, and $\varphi^0_\tau\in[0,\pi]$ and $2\pi-\varphi^0_\tau$ are the two
$\tau$ azimuthal angles that solve the delta function of energy conservation in
Eq.~\eqref{eq:A2}. Their cosines are given by
\begin{align}
\cos\theta_{\tau\pi}&=\frac{2k'^0E_\pi-m_\tau^2-m_\pi^2}{2|\vec k\,'|\,|\vec p_\pi|},\label{eq:costetataupi}\\
\cos\varphi^0_\tau&=\frac{\cos\theta_{\tau\pi}-\cos\theta_\tau\cos\theta_\pi}{\sin\theta_\tau\sin\theta_\pi}
=\frac{2k'^0E_\pi-m^2_\tau-m^2_\pi-2|\vec k\,'|\,|\vec p_\pi|\cos\theta_\tau\cos\theta_\pi}
{2|\vec k\,'|\,|\vec p_\pi|\sin\theta_\tau\sin\theta_\pi}.\label{eq:cosfitau0}
\end{align}
The first follows from the condition $(k'-p_\pi)^2=p_{\nu_\tau}^2=0$, and the second from
expressing $\cos\theta_{\tau\pi}$ in terms of the polar and azimuthal angles of the $\tau$.
As can be seen from Eq.~\eqref{eq:sA}, the information contained in the $TT$ component is lost. In the present case, where the reaction occurs on a neutron bound inside the nucleus rather than on a free neutron, the polar angle of the tau lepton is not kinematically
fixed. Consequently, the integration over $\cos\theta_\tau$ cannot be carried out
using Dirac delta functions. Moreover, the nuclear response functions
$F^{A_Z}_{\nu_\tau \Lambda_c}(q)$ and
$F^{A_Z}_{\nu_\tau \Lambda_c}(q)\times{\cal P}_{A_Z;L,T}(q)$ depend on both $q^2$
and $q^0$ (see below). Unlike the free-neutron case, they cannot be expressed as a
Dirac delta function, $\delta\!\left(2M_nq^0-q^2+M^2_{\Lambda_c}-M_n^2\right)$,
multiplied by terms that depend only on $q^2$.

Note that, upon integrating Eq.~\eqref{eq:sA} over the pion variables, the contributions
that depend on the polarization components vanish,\footnote{This is achieved  by noting that  the  $H[1-|\cos\varphi^0_\pi|]$
function limits $\cos\theta_\pi\in[\cos(\theta_{\tau\pi}+\theta_\tau),
\cos(\theta_{\tau\pi}-\theta_\tau)]$ and that $\sin\theta_\tau\sin\theta_\pi|\sin\varphi^0_\tau|
=\sqrt{(\cos\theta_\pi-\cos(\theta_{\tau\pi}+\theta_\tau))
(\cos(\theta_{\tau\pi}-\theta_\tau)-\cos\theta_\pi)}$, which allows us to perform the integration over $\cos\theta_\pi$ using Eq.~(F.6) of Ref.~\cite{Penalva:2021wye}. This integration eliminates the contribution proportional to ${\cal P}_{A_Z;T}$. Finally, the integration over the pion energy is straightforward, as the integrand depends at most linearly on $E_\pi$. For a given $\tau$ energy $k^{\prime 0}$, the integration limits are $E^-_\pi(k^{\prime 0})$ and $E^+_\pi(k^{\prime 0})$, where
\be
E^{\pm}_\pi(k^{\prime0})=\frac{(m^2_\tau+m^2_\pi)k^{\prime0}\pm
(m^2_\tau-m^2_\pi)\left[(k^{\prime0})^2-m^2_\tau\right]^\frac12}{2m^2_\tau}, \label{eq:epifcnEprima}
\ee
As a result of this integration, the contribution proportional to ${\cal P}_{A_Z;L}$ vanishes.
\label{foot:varfi0}}
and one gets
\be
\int d(\cos\theta_\pi)\,dE_\pi \left(\frac{d^2\sigma_{A_Z}}{d\cos\theta_\pi\,dE_\pi}\right)={\cal B}(\tau\to\pi\nu_\tau)\sigma_{A_Z}
\label{eq:totcs}
\ee
with $\sigma_{A_Z}$ the total nuclear  cross section for tau-neutrino induced 
$\Lambda_c^+$ production
\be
\sigma_{A_Z}=\frac{ G_F^2|V_{cd}|^2}
{2\pi|\vec k\,|}
\int_{m_\tau}^{|\vec k\,|+\Delta M-M_{\Lambda_c}} dE_\tau \sqrt{E_\tau^2-m_\tau^2}\int_{-1}^{+1} d(\cos\theta_\tau)F^{A_Z}_{\nu_\tau \Lambda_c}(q)
\label{eq:secQE}
\ee
where $\Delta M$ is the difference between the mass $M_{A_Z}$ of the initial nucleus and the energy 
of the lowest state of the final $(A-1)Z$ nuclear system. For the $\tau$ kinematics, we consider the 
largest kinematically allowed phase space, assuming that the final nuclear system can absorb any 
momentum transfer. In a realistic calculation, however, the nuclear amplitude $F^{A_Z}_{\nu_\tau 
\Lambda_c}(q)$ determines the kinematically accessible region of phase space.

For SM currents producing a proton instead of a $\Lambda_c^+$, Eq.~\eqref{eq:secQE} reduces to 
Eq.~(2) of Ref.~\cite{Hernandez:2022nmp}, once the different normalization conventions are taken 
into account. In particular, the nuclear amplitude $F_{\nu_\tau}$ introduced in that work is related 
to the present one by $F_{\nu_\tau} = |V_{ud}|^2 F^{A_Z}_{\nu_\tau \Lambda_c}/(2|\vec{k}\,|M_{A_Z})$.
This identification is, however, only formal. Although the two equations have the same structure, our
nuclear amplitude $F_{\nu_\tau}$ differs from that defined in Ref.~\cite{Hernandez:2022nmp}, as the 
latter accounts for the Pauli blocking of the outgoing proton.

Recalling that $E'=\sqrt{M_{\Lambda_c}^2+(\vec{p}-\vec{q},)^2}$ and $q=k'-k$, it is convenient to use
rotational covariance to write
\bea
\int\frac{d^3p}{(2\pi)^3}\int dES_h^{(n)}(E,|\vec p\,|)
\frac{M_nM_{\Lambda_c}}{E\, E'}\,
 \delta(E-E'-q^0)\,E^m &=&a_m, \qquad m=0,1,2 \label{eq:am}\\
\int\frac{d^3p}{(2\pi)^3}\int dES_h^{(n)}(E,|\vec p\,|)
\frac{M_nM_{\Lambda_c}}{E\, E'}\,
 \delta(E-E'-q^0)\,E^mp^j &=&
b_m\, q^j, \qquad m=0,1  \label{aq:bm}\\
\int\frac{d^3p}{(2\pi)^3}\int dES_h^{(n)}(E,|\vec p\,|)
\frac{M_nM_{\Lambda_c}}{E\, E'}\,
 \delta(E-E'-q^0)\,p^jp^k &=&
c_0|\vec q\,|^2\, \delta^{jk}+d_0\,q^jq^k.
\label{eq:c0d0}
\eea
where
\bea
a_m&=&\frac{M_n M_{\Lambda_c}}{4\pi^2|\vec q\,|}\int  d|\vec p\,||\vec p\,|\int
\frac{dE}{E}\,S_h^{(n)}(E,|\vec p\,|) E^m H[1-x_0], \quad m=0,1,2\\
b_m&=&\frac{M_n M_{\Lambda_c}}{8\pi^2|\vec q\,|^3}\int 
d|\vec p\,||\vec p\,| \int
\frac{dE}{E}\,S_h^{(n)}(E,|\vec p\,|) E^m\left(M^2_{\Lambda_c}+|\vec p\,|^2-E^2-q^2+2q^0E\right)H[1-x_0], \quad m=0,1\\
c_0&=&\frac{M_n M_{\Lambda_c}}{8\pi^2|\vec q\,|^3}
\int  d|\vec p\,||\vec p\,|^3\int
\frac{dE}{E}\,S_h^{(n)}(E,|\vec p\,|)(1-x_0^2)H[1-x_0],\\
d_0&=&\frac{M_n M_{\Lambda_c}}{8\pi^2|\vec q\,|^3}
\int  d|\vec p\,||\vec p\,|^3\int
\frac{dE}{E}\,S_h^{(n)}(E,|\vec p\,|)(-1+3x_0^2)H[1-x_0].
\eea
with $x_0=\left|M^2_{\Lambda_c}+|\vec p\,|^2-E^2-q^2+2q^0 E\right|/\left(2|\vec p\,|\,|\vec
q\,|\right)$.
Note that these functions depend on both $q^0$ and $|\vec q,|$ and can be evaluated once the 
neutron-hole SF is specified. We emphasise once again that, for the $d\to u$ transition producing a 
proton rather than a $\Lambda_c^+$ hyperon, the integrations above should, at a minimum, account for
Pauli blocking of the outgoing proton. The nuclear spin-averaged squared amplitude $F^{A_Z}_{\nu\tau
\Lambda_c}(q)$ and the polarisation four-vector ${\cal P}^\mu_{A_Z}$ can be expressed in terms of the
functions defined above, which encode nuclear effects and depend on $\vec{k}$ and $q^\mu$,
\bea
 F^{A_Z}_{\nu_\tau \Lambda_c}(q) &=&
\frac{M^2_n }{2}\Bigg\{
\frac{{\cal B}(\omega)}{M^2_n}\left(a_1|\vec k\,|-b_0\, \vec q\cdot\vec k\right)-{\cal A}(\omega)a_0-
\frac{{\cal C}(\omega)}{M^4_n}\left( |\vec k\,|^2a_2+c_0|\vec q\,|^2|\vec k\,|^2+ d_0\,(\vec q\cdot\vec k)^2-
2b_1|\vec k\,|\,\vec q\cdot\vec k\right) \label{eq:nucl-ampl}
\Bigg\} \\
{\cal P}^\mu_{A_Z}&=& \frac{M^2_n}{ F^{A_Z}_{\nu_\tau \Lambda_c}(q) }\Bigg\{
-\frac{{\cal A_H}(\omega)}{M_n}\Big[t_1^\mu-
\frac{k^{\prime\mu}}{m^2_\tau}\left(a_1 k^{\prime0}-b_0\,\vec q\cdot\vec k\,'\right)
\Big]+\frac{{\cal C_H}(\omega)}{M^3_n}\Big[
t_2^\mu-\frac{k^{\prime\mu}}{m^2_\tau}\left(a_2\, |\vec k\,|k^{\prime0}
-b_1k^{\prime0}\,\vec q\cdot\vec k\right. \no
&-& \left. b_1|\vec k\,|\,\vec q\cdot\vec k\,'
+c_0|\vec q\,|^2\,
\vec k\cdot\vec k\,'+d_0\,(\vec q\cdot\vec k)(\vec q\cdot\vec k\,')\right)\Big]
+\frac{q^\mu_\perp}{M_n}\,\Bigg[
\frac{{\cal D_H}(\omega)}{M^2_n}\left(a_1|\vec k\,|-b_0\, \vec q\cdot\vec k\right)-{\cal B_H}(\omega)a_0 \no
&-&\frac{{\cal E_H}(\omega)}{M^4_n}\left(|\vec k\,|^2a_2+c_0\,|\vec q\,|^2|\vec k\,|^2
+d_0\,(\vec q\cdot\vec k)^2-2b_1|\vec k\,|\,\vec q\cdot\vec k\,\right)\Bigg]
+N^\mu_{TT}\,\frac{|\vec k\times\vec k\,'|}{M^5_n}
\Bigg(M^2_n{\cal F_H}(\omega)\Big[b_0 \left(k^{\prime0}-|\vec k\,|\right)-a_1\Big]\no 
&+&{\cal G_H}(\omega)\left[a_2\,|\vec k\,|-b_1\,
\vec q\cdot\vec
k+c_0|\vec q\,|^2|\vec k\,|+ \left(|\vec k\,|-k^{\prime0}\right)\left(b_1\,|\vec k\,|-d_0\,
\vec q\cdot\vec
k\right)\right]
\Bigg)\Bigg\},\label{eq:polNucl}
\eea
where we have introduced two four-vectors $t_1^\mu = (a_1,b_0\,\vec q\,)$ and
 $t_2^\mu= \left(a_2\, |\vec k\,|-b_1\,\vec q\cdot\vec k,\, b_1|\vec k\,|\vec q-d_0
 (\vec q\cdot\vec k)\vec q-c_0|\vec q\,|^2\,\vec k\,\right)$.

We should point out that in the above equations some 
 struck-nucleon off-shell effects have been neglected. To clarify this point, we will refer
  to the nuclear spin-averaged square amplitude $F^{A_Z}_{\nu_\tau \Lambda_c}(q)$, since 
  the discussion for ${\cal P}^\mu_{A_Z}$ runs in similar terms.  The computation of the 
  nuclear spin-averaged square amplitude requires first the evaluation of the contraction
   $\Big[\sum_{\chi=L,R}\rho_{\chi\chi}\sum_{a,b}W^{ab}_\chi L_{\chi ab}\Big]$, with $W^{ab}_\chi$ the 
   tensor at the hadron level. The former contraction can be expressed in terms of
    $(p\cdot k)$ and $(p\cdot q)$ scalar products times    $q^2$-dependent functions. For $(p\cdot q)$, we have $2(p\cdot q)= 
    (p^2-M_n^2)-(p^{\prime\,2}-M^2_{\Lambda_c}) + M_n^2-M^2_{\Lambda_c}+q^2$, and therefore 
    it reduces to a function of $q^2$, or equivalently of $\omega$, for on-shell hadrons.
    In this evaluation, we have assumed that the initial neutron and the final charmed hyperon are 
    both on shell as in the free case or, more generally, that the difference between their 
    virtualities satisfies $(p^2-M_n^2)-(p^{\prime\,2}-M_{\Lambda_c}^2)\approx 0$ when the two 
    baryons are embedded in the nuclear medium. Under this approximation, the contraction of the 
    leptonic and hadronic tensors can be expressed as a linear combination of $\omega$-dependent 
    functions multiplying different powers of $(p\cdot k)$. This, in turn, allows the functions 
    ${\cal A}(\omega)$, ${\cal B}(\omega)$, and ${\cal C}(\omega)$ in Eq.~\eqref{eq:nucl-ampl} to 
    be factorised despite the nuclear convolution implicit in Eq.~\eqref{eq:nuctenquasie}.   
    Nevertheless, the hole-SF convolution integrals introduced in Eqs.~(\ref{eq:am})--(\ref{eq:c0d0})
    incorporate important nuclear-medium effects. In particular, they determine the kinematically 
    allowed region in $(q^0,|\vec{q}\,|)$ for the QE process and provide a more realistic evaluation
    of the $(p\cdot k)$ and $(p\cdot k)^2$ contributions to the lepton--hadron tensor contraction.
    This treatment of nuclear effects has not been widely adopted in the literature, yet it is 
    particularly well suited to studying the nuclear polarisation vector ${\cal P}^\mu_{A_Z}$ in the
    presence of general dimension-six NP operators. A fully consistent treatment of off-shell 
    effects, however, is not currently available. Indeed, even the definition of hadronic current 
    matrix elements in the nuclear medium involves ambiguities, as these are typically constructed 
    using free-space $n$ and $\Lambda_c$ Dirac spinors and their vacuum masses

 From Eq.~\eqref{eq:polNucl}, one can isolate the longitudinal ($L$), transverse ($T$), and 
 transverse--transverse ($TT$) components of the polarisation vector ${\cal P}_{A_Z}$ associated 
 with this reaction.
%
%
\bea
{\cal P}_{A_Z; L}&=&\frac{M_n}{m_\tau F^{A_Z}_{\nu_\tau \Lambda_c}(q)}\Bigg\{{\cal A_H}(\omega)\Big(a_1|\vec k\,'|
-b_0\frac{k^{\prime0}}{|\vec k\,'|}\,\vec q\cdot\vec k\,'\Big)
-\frac{{\cal C_H}(\omega)}{M^2_n}\Big[
\left(a_2\, |\vec k\,|-b_1\,\vec q\cdot\vec k\,\right)|\vec k\,'|\no 
&-&
\frac{k^{\prime0}}{|\vec k\,'|}\Big((b_1|\vec k\,|-d_0\,\vec q\cdot\vec
k\,)\,
\vec q\cdot\vec k\,'-c_0|\vec q\,|^2\,\vec k\cdot\vec k\,'\Big)\Big]
+\Big(|\vec k||\vec k\,'|-\frac{k^{\prime0}}{|\vec k\,'|}\,\vec k
\cdot\vec k\,'\Big)\Big[-{\cal B_H}(\omega)a_0 \no
&+&
\frac{{\cal D_H}(\omega)}{M^2_n}\left(a_1|\vec k\,|-b_0\, \vec q\cdot\vec k\right)
-\frac{{\cal E_H}(\omega)}{M^4_n}\left(|\vec k\,|^2a_2+c_0\,|\vec q\,|^2|\vec k\,|^2
+d_0\,(\vec q\cdot\vec k)^2-
2b_1|\vec k\,|\,\vec q\cdot\vec k\,\right)\Big]\Bigg\}\\
%
%
{\cal P}_{A_Z;T}&=&\frac{|\vec k\,|\sin\theta_\tau} {F^{A_Z}_{\nu_\tau \Lambda_c}(q)}\Bigg\{-M_n{\cal A_H}(\omega)
b_0+\frac{{\cal C_H}(\omega)}{M_n }\left(
b_1|\vec k\,|-d_0\,\vec q\cdot\vec
k\,+c_0|\vec q\,|^2\right) + \frac{{\cal D_H}(\omega)}{M_n}\left(a_1|\vec k\,|-b_0\, \vec q\cdot\vec k\right)
\no
&-&
M_n{\cal B_H}(\omega)a_0-
\frac{{\cal E_H}(\omega)}{M^3_n}\left(|\vec k\,|^2a_2+c_0\,|\vec q\,|^2|\vec k\,|^2
+d_0\,(\vec q\cdot\vec k)^2-
2b_1|\vec k\,|\,\vec q\cdot\vec k\,\right)\Bigg\}\\
%
%
{\cal P}_{A_Z;TT}&=&\frac{|\vec k\,|\,|\vec k\,'|\sin\theta_\tau}{M_n^3F^{A_Z}_{\nu_\tau \Lambda_c}(q)}
\Bigg\{M^2_n{\cal F_H}(\omega)\Big[b_0 \left(k^{\prime0}-|\vec k\,|\right)-a_1\Big]\no
&+&{\cal G_H}(\omega)\left[a_2\,|\vec k\,|-b_1\,
\vec q\cdot\vec
k+c_0|\vec q\,|^2|\vec k\,|+ \left(|\vec k\,|-k^{\prime0}\right)\left(b_1\,|\vec k\,|-d_0\,
\vec q\cdot\vec
k\right)\right]
\Bigg\}\label{eq:p-projections}
\eea

\subsection{Limits on $E_\pi$ and $\cos\theta_\pi$}
\label{app:limits_integration}
We consider here the integration limits for $E_\pi$ and $\cos\theta_\pi$.
A first estimate can be obtained by neglecting Fermi motion. In this limit,
the kinematically allowed range of the $\tau$ energy and momentum is bounded by
the following minimum and maximum values:
\bea
&&E^{\rm max,min}_\tau=\frac{(|\vec k\,|+M_n)(S+m^2_\tau-M^2_{\Lambda_c})\pm|\vec
k\,|\lambda^{1/2}(S,m^2_\tau,M^2_{\Lambda_c})}{2S}, \ 
|\vec k\,'|_{\rm max,\min}=\sqrt{(E^{\rm max,min}_\tau)^2-m^2_\tau},
\label{eq:etauabsolutemaxmin}
\eea
with $S=M^2_n+2M_n|\vec k\,|$. This automatically fixes the absolute maximum pion energy, 
\bea
E^{\rm max}_\pi=\frac{(m^2_\tau+m^2_\pi)E^{\rm max}_\tau+(m^2_\tau-m^2_\pi)
|\vec k\,'|_{\rm max}}{2m^2_\tau}
\label{eq:epimax}
\eea
As for the minimum pion energy, it is a little more subtle and is given by
\bea
E^{\rm min}_\pi=\left\{\begin{array}{ll}m_\pi&|\vec k\,|\ge\hat k_0\\
E^{\rm int}_\pi&|\vec k\,'|<\hat k_0\end{array}\right.,
\label{eq:epimin}
\eea
where
\bea
\hat k_0&=&\frac{m^2_\tau(M_n-m_\pi)+M_nm^2_\pi+m_\pi(M^{2}_{\Lambda_c}-M^2_n)}
{2(M_n-m_\pi)m_\pi},\label{eq:hatk0}\\
E^{\rm int}_\pi&=&\frac{(m^2_\tau+m^2_\pi)E^{\rm max}_\tau-(m^2_\tau-m^2_\pi)
|\vec k\,'|_{\rm max}}{2m^2_\tau},
\eea
Note that for the pion to be at rest,  the
energy of the intermediate $\tau$ has to be $(m^2_\tau+m^2_\pi)/(2m_\pi)$. However,
for this to happen one needs that $|\vec k\,|\ge \hat k_0$, a bound obtained for forward
$\tau$ production.
For $\cos\theta_\pi$ there is no constraint and its absolute limits
are\footnote{For $\cos\theta_\pi$ to reach $-1$ one needs that
\be
E_\tau\ge |\vec k\,'|\frac{m^2_\tau+m^2_\pi}{m^2_\tau-m^2_\pi}\Longrightarrow
E_\tau\le\frac{m^2_\tau+m^2_\pi}{2m_\pi}.
\ee
Thus, $\cos\theta_\pi=-1$ would be excluded only if 
\be
E^{\rm min}_\tau>\frac{m^2_\tau+m^2_\pi}{2m_\pi}
\ee
which never happens.} $\cos\theta_\pi\in[-1,1]$.
We emphasise that these are only rough estimates, obtained by neglecting Fermi motion. The actual 
kinematically allowed ranges of $E_\pi$ and $\cos\theta_\pi$ are, of course, modified by Fermi motion
and depend on the hole SF adopted to describe the nuclear ground state. \\
Once the pion energy is known, which in this approximation will be comprised between $E_\pi^{\rm max}$
and  $E_\pi^{\rm min}$ given in Eqs.~\eqref{eq:epimax} and \eqref{eq:epimin},  we can put  limits on
the tau energy from the constraint $|\cos\theta_{\tau\pi}|\le1$. One gets
\bea
k^{\prime0}\in[E^{-}_\tau(E_\pi),E^{+}_\tau(E_\pi)],\ 
E^{\pm}_\tau(E_\pi)=\frac{E_\pi(m^2_\tau+m^2_\pi)\pm
|\vec p_\pi|(m^2_\tau-m^2_\pi)}{2 m^2_\pi}
\label{eq:etaumaxmin}
\eea
Here $E^{\pm}_\tau(E_\pi)$ are the maximum and minimum tau energies that can give rise to 
a pion of energy $E_\pi$. However, this interval is not necessarily fully contained in the 
absolute region of tau energies $[E^{\rm min}_\tau,E^{\rm max}_\tau] $  determined by the 
incoming neutrino energy (see Eq.~\eqref{eq:etauabsolutemaxmin}). This is the reason why 
the true $k^{\prime0}$ limits  for a given pion energy are in fact
\bea
k^{\prime0}\in[E^{\rm inf}_\tau,E^{\rm sup}_\tau]
\label{eq:kprime0actuallimits}
\eea
with
\bea
E^{\rm inf}_\tau&=&\left\{\begin{array}{ll}E^{-}_\tau(E_\pi)&
{\rm if}\ E_\pi\notin[E^{-}_\pi(E^{\rm min}_\tau),E^{+}_\pi(E^{\rm min}_\tau)]\\
E^{\rm min}_\tau&{\rm if}\ E_\pi\in[E^{-}_\pi(E^{\rm min}_\tau),
E^{+}_\pi(E^{\rm min}_\tau)]\end{array}\right.,\no
E^{\rm sup}_\tau&=&\left\{\begin{array}{ll}E^{\rm max}_\tau&
{\rm if}\ |\vec k\,|\le\hat k_0\\
\theta(E^{\rm int}_\pi-E_\pi)E^{+}_\tau(E_\pi)+
\theta(E_\pi-E^{\rm int}_\pi)E^{\rm max}_\tau&{\rm if}\ 
|\vec k\,|>\hat k_0\end{array}\right.,
\label{eq:kprime0actuallimits2}
\eea
and the maximum and minimum pion energy for a given tau energy reads
\bea
E_\pi^{\pm}(k^{\prime0})=\frac{(m^2_\tau+m^2_\pi)k^{\prime0}\pm (m^2_\tau-m^2_\pi)
|\vec k\,'|}{2m^2_\tau}
\label{eq:epimaxmintau}
\eea
Moreover, the  $H[1-|\cos\varphi^0_\pi|]$
function can be used to get the limiting values of $\cos\theta_\tau$. One 
has precisely  that
\be
\cos\theta_\tau\in[\cos(\theta_{\tau\pi}+\theta_\pi),
\cos(\theta_{\tau\pi}-\theta_\pi)].
\label{eq:costetataumaxmin}
\ee
Thus, for allowed $E_\pi,\cos\theta_\pi
$ values, we get
\bea
\frac{d^2\sigma_{A_Z}}{d\cos\theta_\pi\,dE_\pi}&=&
\frac{{\cal B}(\tau\to\pi\nu_\tau) m^2_\tau G_F^2|V_{cd}|^2}
{2\pi^2|\vec k\,|(m^2_\tau-m^2_\pi)}
\int_{E^{\rm inf}_\tau}^{E^{\rm sup}_\tau}dk^{\prime0}\int_{\cos(\theta_{\tau\pi}+\theta_\pi)}^{\cos(\theta_{\tau\pi}-\theta_\pi)}
\frac{d(\cos\theta_\tau)F^{A_Z}_{\nu_\tau \Lambda_c}(q)}{
\sqrt{(\cos\theta_\tau-\cos(\theta_{\tau\pi}+\theta_\pi))
(\cos(\theta_{\tau\pi}-\theta_\pi)-\cos\theta_\tau)}}\no
& \times &\Big[1+\frac{2m_\tau}{m^2_\tau-m^2_\pi}
\Big(\frac{{\cal P}_{A_Z;L}}{m_\tau}(E_\pi |\vec k\,'|-|\vec
p_\pi|k^{\prime0}\cos\theta_{\tau\pi})+
\frac{{\cal P}_{A_Z;T}|\vec p_\pi|}{\sin\theta_\tau}(\cos\theta_\pi-\cos\theta_\tau
\cos\theta_{\tau\pi})\Big)\Big]. \label{eq:seccA}
\eea
where we have also made use of the relation
\bea
\sin\theta_\tau\sin\theta_\pi|\sin\varphi^0_\tau|
=\sqrt{(\cos\theta_\tau-\cos(\theta_{\tau\pi}+\theta_\pi))
(\cos(\theta_{\tau\pi}-\theta_\pi)-\cos\theta_\tau)}.
\eea
In the case of SM currents producing a proton instead of a $\Lambda_c^+$, 
Eq.~\eqref{eq:seccA} reduces to Eq.~(10) of Ref.~\cite{Hernandez:2022nmp} with the identifications 
and caveats discussed above in the context of Eq.~\eqref{eq:secQE}. 

Moreover, since $q^2=m_\tau^2-2|\vec k\,|(k^{\prime0}-|\vec k\,'|\cos\theta_\tau)$ and
$q^0=k^0-k^{\prime0}$, Eq.~\eqref{eq:sigmaFinal} follows immediately.

\subsection{Specific results for the LFG model}
\label{app:LFG}

For the particular hole SF of Eq.~\eqref{eq:ShLFG}, one obtains
\be
a^{\rm LFG}_m =\frac{2M_n M_{\Lambda_c}}{(m+1)\pi|\vec q\,|}\int r^2dr\, H\left [E_F(r)-\tilde E\right]\left(E^{m+1}_F(r)-\tilde
E^{m+1}\right), \quad m=0,1,2 \label{eq:aLFGm}
\ee
where we have used that in the LFG case, neglecting off-shell effects, 
\bea
1-x_0^2&=&-\frac{q^2}{|\vec q\,|^2|\vec p\,|^2}\left(E-{\cal E}_+\right)\left(E-{\cal E}_-\right) \label{eq:x0}\\
{\cal E}_{\pm}&=&\frac{q^0}{2}\left(1-\frac{M^2_{\Lambda_c}-M^2_n}{q^2}\right)
\pm\frac{|\vec q\,|}{2}\sqrt{\left(1-\frac{M^2_{\Lambda_c}-M^2_n}{q^2}\right)^2-\frac{4M^2_n}{q^2}}
\eea
where ${\cal E}_-< 0$, since $q^2<0$, and  $0 < {\cal E}_+$ is the minimum neutron energy that 
guarantees  that $x_0\le 1$. In addition in Eq.~\eqref{eq:aLFGm}, $E_F(r)=\sqrt{M^2_n+k^2_F(r)}$ and 
$\tilde E= \max\{M_n,{\cal E}_+\}$.
On the other hand, since the neutron is on-shell, we have that  $q^0 E-\vec p\cdot\vec q=(M^2_n+q^2-
M^2_{\Lambda_c})/2$, from where it follows
\be
b^{\rm LFG}_{m-1}=a^{\rm LFG}_m \frac{q^0}{|\vec
q\,|^2 }-\frac{a^{\rm LFG}_{m-1}}{2|\vec
q\,|^2 }\left(M^2_n+q^2-M^2_{\Lambda_c}\right), \qquad m=1,2
\ee
and from Eqs.~\eqref{eq:aLFGm} and \eqref{eq:x0}, we obtain
\bea
c^{\rm LFG}_0 &=& -\frac{1}{2|\vec
q\,|^4}\left\{q^2a^{\rm LFG}_2 - q^0 a^{\rm LFG}_1  \left(q^2-M^2_{\Lambda_c}+M^2_n\right) +\frac{a^{\rm LFG}_0}{4}\left[ \left(q^2-M^2_{\Lambda_c}+M^2_n\right)^2+ 4M_n^2|\vec
q\,|^2\right]  \right\}\\
d^{\rm LFG}_0 &=& -3c^{\rm LFG}_0  + \frac{a^{\rm LFG}_2-M^2_na^{\rm LFG}_0}{|\vec
q\,|^2}.\label{eq:dLFGm}
\eea
As for the $E_\pi$ and $\cos\theta_\pi$ limits, one has in this case that $E_\tau^{\rm max,min}$ 
are now  given by
\bea
E_\tau^{\rm max,min}=\frac{(|\vec k\,|+E_F)(S_F+m^2_\tau-M^2_{\Lambda_c})\pm (|\vec k\,|- K_F)
\lambda^{1/2}(S_F,m^2_\tau,M^2_{\Lambda_c})}{2S_F},
\label{eq:etaumaxminef}
\eea
with $K_F$, $E_F=\sqrt{M^2_n+K^2_F}$ the maximum  available Fermi momentum and 
energy, and $S_F=(|\vec k\,|+E_F)^2-(|\vec k\,|- K_F)^2=M_n^2+2|\vec k\,|(E_F+K_F)$. 
For the reaction to take place we need that $S_F\ge(m_\tau+M_{\Lambda_c})^2$ which sets the
threshold for the incoming neutrino energy at around $6.3$\,GeV for $K_F\approx 0.27\,$GeV. Note 
that  $\hat k_0$  will also be modified  due to Fermi motion, 

\be
\hat k_0(K_F)=\frac{m^2_\tau(E_F-m_\pi)+E_F m^2_\pi+m_\pi(M^{2}_{\Lambda_c}-M^2_n)+K_F(m^2_\tau-m^2_\pi)}
{2(E_F+K_F-m_\pi)m_\pi}
\label{eq:hatk0ef}
\ee
going from its value of $\approx14.1\,$GeV  in Eq.~(\ref{eq:hatk0}) down to $\approx13.3\,$GeV, 
again for    $K_F\approx 0.27\,$GeV. These new values of  $E^{\rm max,min}_\tau$ and $\hat k_0$ 
determine the absolute interval for  $E_\pi$ according to the discussion in the previous section.\\
The limits for $\cos\theta_\pi$ are however not affected and it can takeany value in the $[-1,1]$ 
interval. Now, for given $E_\pi$ and $\cos\theta_\pi$ values, $E_\tau$ and $\cos\theta_\tau$, or
equivalently $q_0$ and $q^2$, are limited by Eqs.~(\ref{eq:kprime0actuallimits})-(\ref{eq:costetataumaxmin}) 
but using the new values of $E_\tau^{\rm max,min}$ in Eq.~\eqref{eq:etaumaxminef} and $\hat k_0$ in
Eq.~\eqref{eq:hatk0ef}.

\section{Details of the CP-asymmetry determination}
\label{app:asy}
We start from the following expression for $\sigma_{A_Z}$, which is equivalent to Eq.~\eqref{eq:sigaz} but is obtained by integrating over $d^3p,dE$ instead of $d^3p'$, using the energy-conservation delta function $\delta(E-E'-q^0)$ to perform the $E$ integral,
\bea
\sigma_{A_Z}
&=&\frac{ {\cal B}(\tau\to\pi\nu_\tau)  M_{\Lambda_c}m^2_\tau G_F^2|V_{cd}|^2}
{64\pi^6|\vec k\,|(m^2_\tau-m^2_\pi)}
\int\frac{d^3k'}{k^{\prime0}}\int\frac{d^3p_\pi}{ E_\pi\,|\vec k\,'-\vec p_\pi|}
\,\delta(k^{\prime0}-E_\pi-|\vec k\,'-\vec p_\pi|)
\no
& \times &\int \frac{d^3p'}{ E'}
S_h^{(n)}(E'+q^0,|\vec p\,'+\vec q\,|)
\frac{M_n}{E'+q^0}\,
\Big[\sum_{\chi=L,R}\rho_{\chi\chi}\sum_{a,b}W^{ab}_\chi L_{\chi ab}\Big]
\Big(1+\frac{2m_\tau}{m^2_\tau-m^2_\pi}\,p_\pi\cdot{\cal P}
\Big),
\eea

The components ${\cal P}_{L,T,TT}$ defined in Eq.~\eqref{eq:PLTTT} are given by
\bea
{\cal P}_{L}&=&-\frac1{\widetilde{\cal N}(\omega,k\cdot p)}\Big[
\frac{\widetilde{\cal N}_{{\cal H}1}(\omega,k\cdot p)}{M_n}
\Big(\frac{(E'+q^0)|\vec k\,'|}{m_\tau}-\frac{k^{\prime0}}{m_\tau |\vec k\,'|}
(\vec p\,'+\vec k\,'-\vec k\,)\cdot\vec k\,'
\Big)\no
&&+\frac{\widetilde{\cal N}_{{\cal H}2}(\omega,k\cdot p)}{M_n}
\frac{1}{ m_\tau}\Big(-|\vec k\,||\vec k\,'|+\frac{k^{\prime0}}{|\vec k\,'|}
\vec k\cdot\vec k\,'\Big)+\frac{\widetilde{\cal N}_{{\cal H}3}(\omega,k\cdot p)}{M^3_n}
\frac{ m_\tau}{|\vec k\,'|}\,\vec p\,'\cdot(\vec k\,'\times\vec k\,)\Big],\no
{\cal P}_{T}&=&-\frac1{\widetilde{\cal N}(\omega,k\cdot p)}\Big[
\frac{\widetilde{\cal N}_{{\cal H}1}(\omega,k\cdot p)}{M_n}
\Big((\vec k-\vec p\,')\cdot \vec N_{T}
\Big)-\frac{\widetilde{\cal N}_{{\cal H}2}(\omega,k\cdot p)}{M_n}
\frac{(\vec k\times\vec k\,')^2}{|(\vec k\times\vec k\,')\times\vec k\,'|}\no
&&+\frac{\widetilde{\cal N}_{{\cal H}3}(\omega,k\cdot p)}{M^3_n}
\,\vec N_T\cdot\Big((k^{\prime0}\vec k-k^0\vec k\,')\times \vec p\,'\Big)\Big],\no
{\cal P}_{TT}&=&-\frac1{\widetilde{\cal N}(\omega,k\cdot p)}\Big[
\frac{\widetilde{\cal N}_{{\cal H}1}(\omega,k\cdot p)}{M_n}
\Big(-\vec p\,'\cdot \vec N_{TT}
\Big)\no
&&+\frac{\widetilde{\cal N}_{{\cal H}3}(\omega,k\cdot p)}{M^3_n}
\Big(k^{\prime0}(\vec N_{TT}\times\vec k\,)\cdot\vec p\,'
-k^{0}(\vec N_{TT}\times\vec k\,')\cdot\vec p\,'-E'|\vec k\times\vec
k\,'|\Big)\Big].
\eea
In addition, $k\cdot p=|\vec k\,|(E'+k^{\prime0}-
|\vec p\,'|\cos\theta'-|\vec k\,'|\cos\theta_\tau)$.
Note that for nucleons perfectly at rest one has that $\vec p\,'=\vec k-\vec k\,'$ and thus the terms
in $\widetilde{\cal N}_{{\cal H}3}(\omega,k\cdot p)$ in ${\cal P}_L,{\cal P}_T$ and the term in
$\widetilde{\cal N}_{{\cal H}1}(\omega,k\cdot p)$ in ${\cal P}_{TT}$ would cancel identically. This
is in agreement with our findings in Ref.~\cite{Hernandez:2026svn} where the possible CP-violating 
NP effects were limited to ${\cal P}_{TT}$. In the present case, deviations from that situation are 
expected to be small for high energy incident $\nu_\tau$'s. 

Now, for each value of $\vec p\,'$, we define $R'$ as a rotation about the $Z$ axis such that
\bea
R' p'=
\left(E', -|\vec p\,'|\sin\theta', 0,
|\vec p\,'|\cos\theta'\right),
\eea
that we shall keep calling  $p'$ for convenience.
We now exploit rotational invariance and perform the change of integration
variables $R'\vec k\,' \to \vec k\,', R'\vec p_\pi\to \vec p_\pi$, to define the 
asymmetry as
\bea
\Delta\sigma_{A_Z}
&=&\frac{ {\cal B}(\tau\to\pi\nu_\tau)M_{\Lambda_c} m^2_\tau G_F^2|V_{cd}|^2}
{32\pi^5|\vec k\,|(m^2_\tau-m^2_\pi)}
\int\frac{d^3k'}{k^{\prime0}}\int d(\cos\theta_\pi)\int dE_\pi |\vec p_\pi|\Big(
\int_{0}^\pi d\varphi_\pi-\int_\pi^{2\pi}d\varphi_\pi\Big)\frac{\delta(k^{\prime0}-E_\pi-|\vec k\,'-\vec p_\pi|)}{|\vec k\,'-\vec p_\pi|}
\no
&\times&\int d(\cos\theta')\int dE' |\vec p\,'|
S_h^{(n)}(E'+q^0,|\vec p\,'+\vec q\,|)
\frac{M_n}{E'+q^0}\,
\Big[\sum_{\chi=L,R}\rho_{\chi\chi}\sum_{a,b}W^{ab}_\chi L_{\chi ab}\Big]
\Big(1+\frac{2m_\tau}{m^2_\tau-m^2_\pi}\,p_\pi\cdot{\cal P}
\Big)
\eea
and where the integration over the outgoing-hyperon azimuthal angle trivially yields a factor of \(2\pi\). In practice, the above procedure is equivalent to defining, event by event, the positive $Y$ axis along $\vec p\,'\times\vec k$, with the positive $Z$ axis along $\vec k$. The asymmetry is then given by the difference between the cross sections for pions emitted above the $OXZ$ plane ($\varphi_\pi\in[0,\pi]$) and below it ($\varphi_\pi\in[\pi,2\pi]$).

To integrate $\delta(k^{\prime0}-E_\pi-|\vec k\,'-\vec p_\pi|)$ let us note that for an arbitrary rotation $R$
\bea
|\vec k\,'-\vec p_\pi|=| R\vec k\,'- R\vec p_\pi|,\ \ p_\pi\cdot{\cal P}= Rp_\pi\cdot  R{\cal P} .
\eea
Let us consider a rotation $\tilde R$ around the $Z$ axis such that $\tilde R p_\pi=(E_\pi,|\vec p_\pi|\sin\theta_\pi,0,|\vec p_\pi|\cos\theta_\pi)=\tilde p_\pi$
and make the extra change of variables $\vec k\,'\to\tilde R^{-1}\vec k\,'$. We will
obtain
\bea
\Delta\sigma_{A_Z}
&=&\frac{ {\cal B}(\tau\to\pi\nu_\tau)M_{\Lambda_c} m^2_\tau G_F^2|V_{cd}|^2}
{32\pi^5|\vec k\,|(m^2_\tau-m^2_\pi)}
\int\frac{d^3k'}{k^{\prime0}}
\int d(\cos\theta_\pi)\int dE_\pi  |\vec p_\pi| \Big(
\int_{0}^\pi d\varphi_\pi-\int_\pi^{2\pi}d\varphi_\pi\Big)
\frac{\delta(k^{\prime0}-E_\pi-|\vec k\,'-\vec {\tilde p}_\pi|)}{ |\vec k\,'-\vec {\tilde p}_\pi|}
\no
&\times&\int d(\cos\theta') \int dE'\, |\vec p\,'|
S_h^{(n)}(E'+q^0,|\vec p\,'+\tilde R^{-1}\vec k\,'-\vec k\,|)
\frac{M_n}{E'+q^0}\no
&\times&
\Big[\sum_{\chi=L,R}\rho_{\chi\chi}\sum_{a,b}W^{ab}_\chi L_{\chi ab}\Big]
\Big(1+\frac{2m_\tau}{m^2_\tau-m^2_\pi}\,\tilde p_\pi\cdot \tilde R{\cal P}
(\tilde R^{-1}\vec k\,')
\Big)
\eea
with $R^{-1}\vec k^{\,\prime}=|\vec
k\,'|(\sin\theta_\tau\cos(\varphi_\tau+\varphi_\pi),
\sin\theta_\tau\sin(\varphi_\tau+\varphi_\pi),\cos\theta_\tau)$. The delta function can now be solved as 
\bea
\delta(k^{\prime0}-E_\pi-|\vec k\,'-\vec {\tilde p}_\pi|)=
\sum_{j=1}^2\frac{|\vec k\,'-\vec {\tilde p}_\pi|}{|\vec k\,'|\,|\vec p_\pi|
\sin\theta_\pi\sin\theta_\tau|\sin\varphi_\tau|}\delta(\varphi_\tau-
\varphi^{(0,j)}_\tau)
\eea
with  $\varphi^{(0,1)}_\tau, \,\varphi^{(0,2)}_\tau$  the two angles in $[0,2\pi]$ for which their cosine is given by
Eq.~\eqref{eq:cosfitau0}.
We take $\varphi^{(0,1)}_\tau\in [0,\pi]$ and $\varphi^{(0,2)}_\tau=2\pi-
\varphi^{(0,1)}_\tau$. After integration of the delta one gets
\bea
\Delta\sigma_{A_Z}
&=&\frac{ {\cal B}(\tau\to\pi\nu_\tau)M_{\Lambda_c} m^2_\tau G_F^2|V_{cd}|^2}
{32\pi^5|\vec k\,|(m^2_\tau-m^2_\pi)}
 \int dk^{\prime0}\int d\cos\theta_\tau\Big(\int_0^\pi
 d\varphi_\pi-\int_\pi^{2\pi}d\varphi_\pi\Big)\int_{E^-_\pi(k^{\prime0})}^{E^+_\pi(k^{\prime0})}
dE_\pi\no
&\times& \int_{\cos(\theta^0_{\tau\pi}+\theta_\tau)}^{\cos(\theta^0_{\tau\pi}
-\theta_\tau)} \frac{d\cos\theta_\pi}
{\sqrt{(\cos\theta_\pi-\cos(\theta_{\tau\pi}+\theta_\tau))(\cos(\theta_{\tau\pi}-\theta_\tau)-
\cos\theta_\pi)}}\no
&\times&
\int d\cos\theta' \int dE'\, |\vec p\,'|\frac{M_n}{E'+q^0}
\sum_{j=1}^2S_h^{(n)}(E'+q^0,|\vec p\,'+\tilde R^{-1}\vec k^{\,\prime}_j-\vec k\,|)
\Big[\sum_{\chi=L,R}\rho_{\chi\chi}\sum_{a,b}W^{ab}_\chi L_{\chi ab}\Big]\no
&\times&
\Bigg\{1+\frac{2m_\tau}{m^2_\tau-m^2_\pi}\,\Big[
{\cal P}_L(\tilde R^{-1}\vec k^{\,\prime}_j)\Big(\frac{|\vec
k\,'|}{m_\tau} E_\pi-\frac{k^{\prime0}}{m_\tau}|\vec
p_\pi|\cos\theta_{\tau\pi}\Big)\no
&+&{\cal P}_T(\tilde R^{-1}\vec k^{\,\prime}_j)\frac{|\vec p_\pi|}{\sin\theta_\tau}\Big(\cos\theta_\pi-\cos\theta_{\tau\pi}\cos\theta_\tau\Big)
+{\cal P}_{TT}(\tilde R^{-1}\vec k^{\,\prime}_j)|\vec p_\pi|\Big(\sin\theta_\pi\sin\varphi^{(0,j)}_\tau\Big)\Big]
\Bigg\}.
\eea
where we have made explicit the product $\tilde p_\pi\cdot R{\cal P}(\tilde R^{-1}\vec k^{\,\prime}_j)$ and we have used  the definition of $\cos\theta_{\tau\pi}$ in Eq.~\eqref{eq:costetataupi} and the information contained in the discussion in  footnote~\ref{foot:varfi0}.

The actual values for ${\cal P}_{L,T,TT}(\tilde R^{-1}\vec k^{\,\prime}_j)$ are 
\bea
{\cal P}_{L}(\tilde R^{-1}\vec k^{\,\prime}_j)&=&-\frac1{\widetilde{\cal N}(\omega,k\cdot p)}\Big[
\frac{\widetilde{\cal N}_{{\cal H}1}(\omega,k\cdot p)}{M_n}\Big(\frac{(E'+q^0)|\vec k\,'|}{m_\tau}\no
& - &
\frac{k^{\prime0}}{m_\tau |\vec k\,'|}
\Big[|\vec p\,'|\,|\vec k\,'|(-\sin\theta'\sin\theta_\tau\cos(\varphi^{(0,j)}_\tau+\varphi_\pi)
+\cos\theta'\cos\theta_\tau)
+|\vec k\,'|^2-|\vec k\,|\,|\vec k\,'|\cos\theta_\tau\Big]
\Big)\no
&+&\frac{\widetilde{\cal N}_{{\cal H}2}(\omega,k\cdot p)}{M_n m_\tau}\left(-|\vec k\,||\vec k\,'|+k^{\prime0}|\vec k\,|\cos\theta_\tau\right)-\frac{\widetilde{\cal N}_{{\cal H}3}(\omega,k\cdot p)}{M^3_n}
 m_\tau|\vec p\,'|\,|\vec k\,|\sin\theta'\sin\theta_\tau
 \sin(\varphi^{(0,j)}_\tau+\varphi_\pi)\Big],\nonumber\\
{\cal P}_{T}(\tilde R^{-1}\vec k^{\,\prime}_j)&=&\frac1{\widetilde{\cal N}(\omega,k\cdot p)}\Big[
\frac{\widetilde{\cal N}_{{\cal H}1}(\omega,k\cdot p)}{M_n}
\Big(|\vec k|\sin\theta_\tau-|\vec
p\,'|(\sin\theta'\cos\theta_\tau\cos(\varphi^{(0,j)}_\tau+\varphi_\pi)-\cos\theta'\sin\theta_\tau)
\Big)\no
&+&\frac{\widetilde{\cal N}_{{\cal 
H}2}(\omega,k\cdot p)}{M_n}
|\vec k\,|\sin\theta_\tau-\frac{\widetilde{\cal N}_{{\cal H}3}(\omega,k\cdot p)}{M^3_n}
\Big(k^0|\vec k\,'|\,|\vec p\,'|-k^{\prime0}|\vec k\,|\,|\vec p\,'|\cos\theta_\tau
\Big)\sin\theta'\sin(\varphi^{(0,j)}_\tau+\varphi_\pi)\Big],\nonumber\\
{\cal P}_{TT}(\tilde R^{-1}\vec k^{\,\prime}_j)&=&
\frac1{\widetilde{\cal N}(\omega,k\cdot p)}\Big[
\frac{\widetilde{\cal N}_{{\cal H}1}(\omega,k\cdot p)}{M_n}
|\vec p\,'|\sin\theta'\sin(\varphi^{(0,j)}_\tau+\varphi_\pi)-\frac{\widetilde{\cal N}_{{\cal H}3}(\omega,k\cdot p)}{M^3_n}
\Big(-k^{\prime0}|\vec k\,|\,|\vec
p\,'|\sin\theta'\cos(\varphi^{(0,j)}_\tau+\varphi_\pi)\no
&+&k^{0}|\vec k\,'|\,|\vec p\,'|(\sin\theta'
\cos\theta_\tau\cos(\varphi^{(0,j)}_\tau+\varphi_\pi)+
\cos\theta'\sin\theta_\tau)-E'|\vec k\,|\,|\vec
k\,'|\sin\theta_\tau\Big)\Big].
\eea

To get  the final expression given in the main text we have done the following:
\begin{enumerate}
\item In the integral $\int_\pi^{2\pi} d\varphi_\pi\cdots$ make the change of variable
$2\pi-\varphi_\pi\to\varphi_\pi$ so that it transforms into an integral in the $[0,\pi]$ interval.
\item Next, use $\varphi_\tau^{(0,2)}=2\pi-\varphi_\tau^{(0,1)}$ to eliminate $
\varphi_\tau^{(0,2)}$ in favor of $\varphi_\tau^{(0,1)}$. 
    \item Group all the terms that depend on $\varphi_\tau^{(0,1)}+\varphi_\pi$ together and all that 
depend on $\varphi_\tau^{(0,1)}-\varphi_\pi$ together. The latter are the ones that originate 
from $\varphi_\tau^{(0,2)}+\varphi_\pi$ terms once $\varphi_\tau^{(0,2)}=2\pi-\varphi_\tau^{(0,1)}$ is used.
\item For the first group make the change of variable $\varphi_\tau^{(0,1)}+
 \varphi_\pi+\pi\to\tilde\varphi$ and for the second make the change of variable
$\varphi_\tau^{(0,1)}-\varphi_\pi+\pi\to\tilde\varphi$
\end{enumerate}
%

The final result reads
%
\bea
\Delta\sigma_{A_Z}
&=&\frac{ {\cal B}(\tau\to\pi\nu_\tau)M_{\Lambda_c} m^2_\tau G_F^2|V_{cd}|^2}
{16\pi^5|\vec k\,|(m^2_\tau-m^2_\pi)}
\int dk^{\prime0}\int d\cos\theta_\tau\Bigg[\int_{\varphi^{(0,1)}_\tau+\pi}^{\varphi^{(0,1)}_\tau+2\pi}
 d\tilde\varphi-\int_{\varphi^{(0,1)}_\tau}^{\varphi^{(0,1)}_\tau+\pi} d\tilde\varphi\Bigg]\no
&\times&\int_{E^-_\pi(k^{\prime0})}^{E^+_\pi(k^{\prime0})}
dE_\pi \int_{\cos(\theta_{\tau\pi}+\theta_\tau)}
^{\cos(\theta_{\tau\pi}-\theta_\tau)}  \frac{d\cos\theta_\pi}{
\sqrt{(\cos\theta_\pi-\cos(\theta_{\tau\pi}+\theta_\tau))(\cos(\theta_{\tau\pi}-\theta_\tau)-
\cos\theta_\pi)}}
\no
&\times&\int d\cos\theta' \int dE'\, |\vec p\,'|\frac{1}{E'+q^0}
S_h^{(n)}(E'+q^0,|\vec p\,'+\vec {\tilde k}{\,'}-\vec k\,|)
\Big(-\frac{2m_\tau}{m^2_\tau-m^2_\pi}\Big)\widetilde{\cal N}_{{\cal H}3}(\omega,k\cdot p)\no
&\times&
\Bigg\{
\Big(\frac{|\vec
k\,'|}{m_\tau} E_\pi-\frac{k^{\prime0}}{m_\tau}|\vec
p_\pi|\cos\theta_{\tau\pi}\Big)\Big(
 m_\tau|\vec p\,'|\,|\vec k\,|\sin\theta'\sin\theta_\tau
 \sin\tilde\varphi
\Big)\no
&-&\frac{|\vec p_\pi|}{\sin\theta_\tau}\Big(\cos\theta_\pi-\cos\theta_{\tau\pi}\cos\theta_\tau\Big)\left[
\left(k^0|\vec k\,'|\,|\vec p\,'|-k^{\prime0}|\vec k\,|\,|\vec p\,'|\cos\theta_\tau
\right)\sin\theta'\sin\tilde\varphi
\right]\no
&+&|\vec p_\pi|\sin\theta_\pi\sin\varphi^{(0,1)}_\tau
\Big(k^{\prime0}|\vec k\,|\,|\vec
p\,'|\sin\theta'\cos\tilde\varphi+k^{0}|\vec k\,'|\,|\vec p\,'|
(\cos\theta'\sin\theta_\tau-\sin\theta'
\cos\theta_\tau\cos\tilde\varphi)\no
&-&E'|\vec k\,|\,|\vec
k\,'|\sin\theta_\tau\Big)
\Bigg\},\label{eq:asyfinal}
\eea
where $\vec {\tilde k}\,'=|\vec k\,'|(-\sin\theta_\tau\cos\tilde\varphi,
-\sin\theta_\tau\sin\tilde\varphi,\cos\theta_\tau)$, and $ k\cdot p
=|\vec k\,|(E'+k^{\prime0}- |\vec p\,'|\cos\theta'-|\vec k\,'|\cos\theta_\tau)$.
Besides, $\varphi_\tau^{(0,1)}$ is the angle in the $[0,\pi]$ interval for which its cosine is given 
by Eq.~\eqref{eq:cosfitau0} and $E^\pm_\pi(k^{\prime0})$, given in Eq.~\eqref{eq:epimaxmintau},
are the maximum and minimum pion energy allowed for an intermediate $\tau$ with $(k^{\prime0},\vec 
k\,')$ four-momentum. Within the energy interval determined by $E^\pm_\pi(k^{\prime0})$ it is  
guaranteed  that $|\cos\theta_{\tau\pi}|\le1$.
\subsection{Asymmetry in the LFG approximation}
\label{app:asyLFG}
In the case that we use the  LFG model we have a new delta that we have to solve. For the integration interval that we have, we get
\bea
\delta(E'+q^0-\sqrt{M^2_n+(\vec p\,'+\vec {\tilde k}{\,'}-\vec k\,)^2}\,)
=\frac{E'+q^0}{|\vec k\,'|\,|\vec p\,'|\sin\theta_\tau\sin\theta'
|\sin\tilde\varphi_{0,1}|}
(\delta(\tilde\varphi-\tilde\varphi_{0,1})
+\delta(\tilde\varphi-\tilde\varphi_{0,2}))
\eea
with $\tilde\varphi_{0,1},\,\tilde\varphi_{0,2}$ the two solutions in 
$[\varphi^{(0,1)}_\tau,\,\varphi^{(0,1)}_\tau+2\pi]$  for which their cosine is
\bea
\cos\tilde\varphi_{0}=\frac{M^2_{\Lambda_c}-M^2_n+q^2
+2E'q^0+2|\vec p\,'|\cos\theta'(|\vec k\,|
-|\vec k^{\,\prime}|\cos\theta_\tau)}
{2|\vec k^{\,\prime}|\,|\vec p^{\,\prime}|\sin\theta_\tau\sin\theta'}.
\label{eq:cosfi0}
\eea

Demanding  that  $|\cos\tilde\varphi_{0,1}|\le1$ limits $\cos\theta'$
to the interval $[\cos\theta'_-,\cos\theta'_+]$ with
\small
\bea
\cos\theta'_{\pm}=\frac{-(|\vec k\,|-|\vec k\,'|
\cos\theta_\tau)(M^2_{\Lambda_c}-M^2_n+q^2+2E'q^0)\pm
|\vec k\,'|\sin\theta_\tau
\sqrt{4|\vec p\,'|^2|\vec q\,|^2-(M^2_{\Lambda_c}-M^2_n+q^2+2E'q^0)^2}}{2|\vec p\,'|
|\vec q\,|^2}.
\label{eq:costhetap}
\eea
\normalsize
For this to happen, one needs\footnote{Note this condition also guarantees that 
$|\cos\theta'_{\pm}|\le1$. Indeed, both $\cos\theta'_{\pm}$ values can be written  as the scalar product of 
 dimension-two unit vectors. Those  unit vectors are
\bea
\Big(\frac{-(|\vec k\,|-|\vec k\,'| \cos\theta_\tau)}{|\vec q\,|}, \pm\frac{|\vec k\,'|\sin\theta_\tau}{|\vec q\,|}\Big),\ 
\Big(\frac{M^2_{\Lambda_c}-M^2_n+q^2+2E'q^0}{2|\vec p\,'|\,|\vec q\,|},
\frac{\sqrt{4|\vec p\,'|^2|\vec q\,|^2-(M^2_{\Lambda_c}-M^2_n+q^2+2E'q^0)^2}}{2|\vec p\,'|\,|\vec q\,|}\Big)
\eea
and their scalar product is always $\le1$ in modulus. }
\bea
4|\vec p\,'|^2|\vec q\,|^2>(M^2_{\Lambda_c}-M^2_n+q^2+2E'q^0)^2,
\label{eq:condthetaprime}
\eea
which forces
\bea
E'>\frac{q^0(M^2_{\Lambda_c}-M^2_n+q^2)+|\vec
q\,|\sqrt{(M^2_{\Lambda_c}-M^2_n+q^2)^2-
4q^2M^2_{\Lambda_c}}}{-2q^2}.
\eea
Besides, the presence of the Heaviside function $H(k_F(r)-|\vec p\,'+\vec{\tilde k}{'}_{\!1}-\vec k\,|)\equiv
H(E_F(r)-(E'+q^0)),$ tell us that
\bea
E'<E_F(r)-q^0=E_F(r)+|\vec k\,|-k^{\prime0}.
\label{eq:eplim}
\eea
As for $k^{\prime0}$, its minimum and maximum value are given by Eq.~\eqref{eq:etaumaxminef}.
One can also rewrite
\bea
\sin\theta_\tau\sin\theta'|\sin\tilde\varphi_{0,1}|=\frac{|\vec q\,|}{|\vec k\,'|}\sqrt{(\cos\theta'-\cos\theta^\prime_-)
(\cos\theta^\prime_+-\cos\theta')}.
\eea
Taking $\tilde\varphi_{0,1}<\tilde\varphi_{0,2}$, one arrives at the final expression 
\bea
\Delta\sigma_{A_Z}
&=&\frac{ {\cal B}(\tau\to\pi\nu_\tau)M_{\Lambda_c} m^2_\tau G_F^2|V_{cd}|^2}
{\pi^4|\vec k\,|(m^2_\tau-m^2_\pi)}
\int dk^{\prime0}\int d(\cos\theta_\tau)
\int_{E^-_\pi(k^{\prime0})}^{E^+_\pi(k^{\prime0})}
dE_\pi \no
&\times&\int_{\cos(\theta_{\tau\pi}+\theta_\tau)}
^{\cos(\theta_{\tau\pi}-\theta_\tau)}  \frac{d(\cos\theta_\pi)}{
\sqrt{(\cos\theta_\pi-\cos(\theta_{\tau\pi}+\theta_\tau))(\cos(\theta_{\tau\pi}-\theta_\tau)-
\cos\theta_\pi)}}
\no
&\times&\int dr\, r^2\int dE'
\, H(E_F(r)-(E'+q^0))\int_{\cos\theta^\prime_-}^{\cos\theta^\prime_+}
\frac{ d(\cos\theta') }{|\vec q\,|
\sqrt{(\cos\theta'-\cos\theta^\prime_-)
(\cos\theta^\prime_+-\cos\theta')}
}\no
&\times&\Big(-\frac{2m_\tau}{m^2_\tau-m^2_\pi}\Big)
\widetilde{\cal N}_{{\cal H}3}(\omega,k\cdot p)
\Bigg\{\Bigg[
\Big(\frac{|\vec
k\,'|}{m_\tau} E_\pi-\frac{k^{\prime0}}{m_\tau}|\vec
p_\pi|\,\cos\theta_{\tau\pi}\Big)\Big(
 m_\tau|\vec p\,'|\,|\vec k\,|\sin\theta'\sin\theta_\tau
 \Big)\no
&-&\frac{|\vec p_\pi|}{\sin\theta_\tau}\Big(\cos\theta_\pi-\cos\theta_\tau\cos\theta_{\tau\pi}\Big)
\Big(-k^{\prime0}|\vec k\,|\,|\vec p\,'|\cos\theta_\tau
+k^0|\vec k\,'|\,|\vec p\,'|
\Big)\sin\theta'\Bigg](-\sin\tilde\varphi_{0,1})H\left(\xi\right)\no
&+&
H\left(-\xi\right){\rm sign}
[\cos\tilde\varphi_{0,1}]
 |\vec p_\pi|\sin\theta_\pi\sin\varphi^{(0,1)}_\tau 
\Big(k^{\prime0}|\vec k\,|\,|\vec
p\,'|\sin\theta'\cos\tilde\varphi_{(0,1)}\no
&+&k^{0}|\vec k\,'|\,|\vec p\,'|(-\sin\theta'
\cos\theta_\tau\cos\tilde\varphi_{(0,1)}+
\cos\theta'\sin\theta_\tau)-E'|\vec k\,|\,|\vec
k\,'|\sin\theta_\tau\Big)
\Bigg\},\label{eq:assyLFG}
\eea
with $E_F(r)$ denoting the local Fermi energy, $\xi=|\cos\varphi^{0}_\tau|-|\cos\tilde\varphi_{0,1}|$, where $\cos\varphi_\tau^0$ is given in Eq.~\eqref{eq:cosfitau0}, and $\tilde\varphi_{0,1}$ is the smaller of the two angles within the interval $[\varphi_{\tau}^{(0,1)},\varphi_{\tau}^{(0,1)}+2\pi]$ whose cosine is given by Eq.~\eqref{eq:cosfi0}. Furthermore, $\cos\theta'_{\pm}$ (Eq.~\eqref{eq:costhetap}) denote the maximum and minimum allowed values of $\cos\theta'$ for a given intermediate-$\tau$ three-momentum $\vec k\,'$ and final hyperon energy $E'$.

\bibliography{LCTAU}
\end{document}